\documentclass[12pt]{article}
\pdfoutput=1

\usepackage[bulletsep]{collref}
\usepackage{amssymb,graphicx}
\usepackage[intlimits]{amsmath}
\usepackage{bbm}
\usepackage[small]{subfigure}

\usepackage{MnSymbol}

\usepackage{arydshln}

\usepackage{xcolor}

\newcommand{\cl}[1]{\mathsf{ #1}}

\makeatletter \@addtoreset{equation}{section} \makeatother

\makeatletter
\let\old@startsection=\@startsection
\let\oldl@section=\l@section
\renewcommand{\@startsection}[6]{\old@startsection{#1}{#2}{#3}{#4}{#5}{#6\mathversion{bold}}}
\renewcommand{\l@section}[2]{\oldl@section{\mathversion{bold}#1}{#2}}
\makeatother

\makeatletter
\let\old@makecaption=\@makecaption
\def\@makecaption{\small\old@makecaption}
\makeatother

\renewcommand{\geq}{\geqslant}
\newcommand{\Bst}{{\rm{Bst}}}
\newcommand{\an}{\alpha ^{(n)}}
\newcommand{\ao}{\alpha ^{(0)}}
\newcommand{\al}{\alpha ^{(1)}}
\newcommand{\ha}{\hat{\alpha}}

\newcommand{\EE}{{\bf E}}
\newcommand{\KK}{{\bf K}}

\begin{document}

%%%%%%%%%%%%%%%%%%%%%%%%%%%%%%%%%%%%%%%%%%%%%%%%%%%%%%%%%%%%%%%%%%%%%%%%%%%%%%%%
\thispagestyle{empty}

\renewcommand{\thefootnote}{\fnsymbol{footnote}}
\setcounter{footnote}{0}

\begin{center}
{\Large\textbf{\mathversion{bold} 't~Hooft loops
 in N=4 super-Yang-Mills}
\par}

\vspace{0.8cm}

\textrm{Charlotte~Kristjansen$^{1}$ and
Konstantin~Zarembo$^{1,2}$}
\vspace{4mm}

\textit{${}^1$Niels Bohr Institute, Copenhagen University, Blegdamsvej 17, \\ 2100 Copenhagen, Denmark}\\
\textit{${}^2$Nordita, KTH Royal Institute of Technology and Stockholm University,
Hannes Alfv\'{e}ns V\"{a}g 12, 114 19 Stockholm, Sweden}\\
\vspace{0.2cm}
\texttt{kristjan@nbi.dk, zarembo@nordita.org}
%\vspace{3mm}

\vspace{3mm}

%%%%%%%%

\par\vspace{1cm}

\textbf{Abstract} \vspace{3mm}

\begin{minipage}{13cm} 

We set up a perturbative framework for the 't~Hooft line in the $\mathcal{N}=4$ super-Yang-Mills theory, and apply it to correlators thereof with Wilson loops and local operators. Using this formalism we obtain a number of perturbative and non-perturbative results that directly connect to localization, holography and integrability.

\end{minipage}
\end{center}

\vspace{0.5cm}

\newpage
\setcounter{page}{1}
\renewcommand{\thefootnote}{\arabic{footnote}}
\setcounter{footnote}{0}

\newcommand{\CC}{\mathbbm{C}}
\newcommand{\kkb}{{\color{teal} \mathring{a}}}
\newcommand{\knb}{{\color{blue} \o}}
\newcommand{\nkb}{{\color{blue} \o}}
\newcommand{\nnb}{{\color{violet} \ae}}
\newcommand{\Yvec}{{\bf Y}_{J\ell M}^{(q)}(\theta,\phi)}
\newcommand{\cpo}{{\rm CPO}}
\newcommand{\bl}[1]{{\color{blue} #1}}

%%%%%%%%%%%%%%%%%%%%%%%%%%%%%%%%%%%%%%%%%%%%%%%%%%%%%%%%%%%%%%%%%%%%%%%%%%%%%%%%

\section{Introduction}

Holography has provided us with a large class of defect conformal field theories of different dimensions for which conformal
data can be explicitly extracted either by perturbative computations or by means of exact methods such as supersymmetric localization, bootstrap or integrability~\cite{Pestun:2016zxk,Liendo:2012hy,Kristjansen:2024dnm}. Whereas defects defined by the insertion of products of local or non-local operators such as Wilson-loops can be treated by means of standard perturbative machinery, see see e.g.~\cite{Erickson:2000af}, defects defined in terms of disorder operators such as certain types of domain walls~\cite{Buhl-Mortensen:2016pxs,Buhl-Mortensen:2016jqo,GimenezGrau:2018jyp,Gimenez-Grau:2019fld}, surface defects~\cite{Choi:2024ktc,deLeeuw:2024qki} and monopoles~\cite{Kristjansen:2023ysz} require non-trivial operations to set up the perturbative framework.

We have recently taken the first steps in setting up the perturbative framework for the handling of a Dirac monopole embedded
in ${\cal N}=4$ SYM, and we have demonstrated that the underlying supersymmetry  simplifies computations and makes the defect set-up integrable~\cite{Kristjansen:2023ysz,Gombor:2024api}.
In the present paper we streamline our derivations and generalize them to arbitrary values of the monopole charge.
Furthermore, in view of the recent renewed interest in 
fermion-monopole scattering~\cite{Brennan:2023tae,vanBeest:2023dbu,vanBeest:2023mbs,Khoze:2023kiu} 
we extend our analysis to include the quantization of the fermionic fields in the 
monopole background. 

Having set up the perturbative program we proceed to the computation of respectively one-point functions and the expectation
value of a Wilson line in the monopole background. In both cases our initially perturbative approach leads to insights which
go far beyond perturbation theory.

First, our computations reveal intriguing truncations of the perturbative series for one-point functions of chiral primaries similar to those observed for surface defects in~\cite{Choi:2024ktc}, where these one-point functions  were shown
to be finite polynomials in the 't Hooft coupling and given entirely by Feynman diagrams without internal vertices.
In our case a similar truncation holds for chiral primaries of odd length. For even length, diagrams without internal vertices
account for all contributions up to wrapping length,  whereafter additional wrapping corrections come into play. Revisiting
the D3-D5 domain wall model we find that a similar phenomenon takes place there~\cite{Buhl-Mortensen:2017ind}.

Secondly, our perturbative approach to the computation of a Wilson line expectation value exposes the possibility of
considering a particular double scaling limit where a complete resummation of diagrams to all orders can be performed. This limit
involves scaling the R-symmetry angle between the Wilson and the 't Hooft line, to imaginary infinity much in analogy with the fishnet limit considered in~\cite{Gurdogan:2015csr}.  The resummed expectation value can be compared to the result of a string theory computation and a perfect match is found.

Our paper is organized as follows. In section \ref{background-field-sec} we perform the quantization of  the 't~Hooft line in the background field gauge.  We analyze the kinematics and determine the
propagators of the various fields by means of spectral decomposition.  In the process, we split the bosonic fields into easy and hard fields with the terminology referring to the complexity of their mixing and treat each group separately.  Fermions are likewise treated separately. Infinities are handled by dimensional regularization.
 In section~\ref{one-point} we calculate one-point functions of local operators in the monopole background. We start by considering chiral primaries where we observe the truncations of the perturbative series mentioned above. Subsequently,
we turn to computing the one-loop correction to the one-point function for general, non-protected scalar operators. Finally,
in section~\ref{Wilson}  we compute the correlator of the 't~Hooft line and a parallel supersymmetric Wilson line, first at one-loop
order and secondly by a complete resummation of diagrams to all orders in the above mentioned double scaling limit.  The section culminates with the demonstration of a match with string theory\footnote{The correlator measures the interaction potential between a charge and a monopole, and has been studied holographically in \cite{Minahan:1998xb,Gorsky:2009pc}.}.
 Section~\ref{conclusion} contains our conclusion.

\section{Quantizing the monopole}\label{background-field-sec}

A 't~Hooft loop is a disorder operator characterized by singular behavior of the fields on a given space-time contour \cite{tHooft:1977nqb}. For an infinite 't~Hooft line, the boundary condition is the field of a  magnetic monopole \cite{Kapustin:2005py}:
\begin{align}\label{monopoleF}
 \cl{F}_{ij}&=q\,\varepsilon _{ijk}\,\frac{x^k}{r^3},
\\
\label{monopolePhi}
 \cl{\Phi }_I&=q\,\frac{n_I}{r}\,.
\end{align}
The monopole is an Abelian object residing in a $U(1)$ subgroup embedded in a larger non-Abelian symmetry, for instance as an upper-left corner in the $U(N)$ which  implies the following decomposition of the fields in the adjoint:
\begin{equation}\label{N+k-decomp}
\begin{array} {ll}
   A_\mu ,\Phi _i,\Psi= \left[
 \begin{array}{c:ccc}
 \kkb & \knb & \knb & \knb \\
 \hdashline
  \nkb & \nnb  & \nnb & \nnb \\
   \nkb & \nnb & \nnb & \nnb  \\
    \nkb & \nnb & \nnb & \nnb \\
 \end{array}
 \right].
  \end{array}
\end{equation}
The monopole resides in the $\kkb$ corner,
the $ \knb$ components are charged under the corresponding $U(1)$ and interact with the monopole, while all other components are $U(1)$-neutral.

The 't~Hooft loop also picks an $R$-symmetry direction $n_I$,  set to $n_I=(1,\mathbf{0})$, for definiteness.
The monopole charge $q$ is quantized in half-integers, by a famous argument of Dirac \cite{Dirac:1931kp}\footnote{For a recent discussion from the perspective of generalized symmetries see \cite{Hull:2024uwz}.}. The smallest, elementary magnetic charge $q=1/2$ corresponds to a single $D1$-brane pinned to a $D3$-brane, while higher-charge 't~Hooft loops are composite objects dual to $2q$ coincident D1-branes that can be split apart at no energy cost. We  will nonetheless keep $q$ arbitrary whenever possible.

Correlation functions of the 't~Hooft line can be computed  perturbatively by  the conventional background field method. The  background itself cannot receive quantum corrections. This is because the radial dependence is stipulated by conformal symmetry,  and only overall coefficients can get renormalized, but renormalization of the magnetic charge is forbidden by the Dirac condition, while the scalar is related to the magnetic field by supersymmetry and cannot renormalize independently.  

Our goal is to develop a systematic perturbative framework for the 't~Ho\-oft line in the  $\mathcal{N}=4$ super-Yang-Mills treating (\ref{monopoleF}), (\ref{monopolePhi}) as classical background fields.

\subsection{Background field expansion}\label{BFE-sec}

The $\mathcal{N} =4$ theory is best viewed as a reduction of the 10d  super-Yang-Mills  to four dimensions \cite{Brink:1976bc}, or to $4-2\varepsilon $ dimensions for the purpose of UV regularization. The gauge fields $A_\mu $ and the six (or $6+2\varepsilon $) scalars $\Phi _I$ form a ten-dimensional vector potential: $A_M=(A_\mu ,\Phi _I)$, and the fermions fit into a single Majorana-Weyl spinor of $SO(1,9)$. The basis of 10d Dirac matrices best tailored for the $4+6$ decomposition is constructed from the 4d and 6d Dirac matrices as follows: 
\begin{equation}\label{10D-Dirac}
 \Gamma ^\mu =\gamma ^\mu \otimes \mathbbm{1},
 \qquad 
 \Gamma ^I=\gamma ^5\otimes\hat{\gamma }^I.
\end{equation}
The 10d spinor index then naturally splits into the space-time and R-sym\-metry labels: $\Psi _{\alpha A}$. 

The Lagrangian is
\begin{equation}
 \mathcal{L}=\frac{N}{\lambda }\mathop{\mathrm{tr}}
 \left(
 \frac{1}{2}\,F_{MN}^2+i\bar{\Psi }\Gamma ^MD_M\Psi 
 \right),
\end{equation}
where $\bar{\Psi }=\Psi ^\top C$ and $\Gamma ^{11}\Psi =\Psi $, and we take the metric to be Euclidean. Compatibility of the Majorana and Weyl conditions does not really depend on the metric signature \cite{Stone:2020vva}. The Euclidean Dirac matrices cannot be chosen real in 10d, as often required for Majorana fermions,
but the Euclidean path integral with the Majorna-Weyl condition imposed is well defined irrespective of that \cite{Stone:2020vva} and when dealing with the Lagrangian we always keep in mind its path-integral quatization.
The gauge potentials are Hermitian by convention, with the covariant derivative acting as $D_M=\partial _M-iA^{\rm adj}_M$, $A^{\rm adj}\equiv [A,\cdot ]$. For the R-symmetry directions  $\partial _I=0$.

The standard gauge-fixing choice for  a classical background $\cl{A}_M$ is
\begin{equation}
 \mathcal{L}_{\rm gf}=\frac{N}{\lambda }\,\mathop{\mathrm{tr}}\left(
 \cl{D}_MA^M
 \right)^2,
\end{equation}
where $\cl{D}_M$ is the background covariant derivative: $\cl{D}_M=\partial _M-i\cl{A}_M^{\rm adj}$. The ghost action in this gauge is
\begin{equation}
 \mathcal{L}_{\rm gh}=\frac{2N}{\lambda }\,\mathop{\mathrm{tr}}
 \cl{D}_M\bar{c}D^Mc.
\end{equation}

The full quantum fields $A_M$ and $\Psi $ are then split into the classical part and fluctuations:
\begin{equation}\label{backgr-exp}
 A_M=\cl{A}_M+a_M,\qquad \Psi =\psi.
\end{equation}
When expanded around the classical solution the gauge-fixed Lagrangian becomes:
\begin{align}\label{Ltot}
 \mathcal{L}_{\rm tot}=&\frac{1}{g^2}\mathop{\mathrm{tr}}\left\{
 a^M\left(-\cl{D}^2\eta _{MN}+2i\cl{F}_{MN}^{\rm adj}\right)a^N
 +i\bar{\psi }\Gamma ^M\cl{D}_M\psi +2\bar{c}\left(-\cl{D}^2\right)c
 \vphantom{\frac{1}{2}}
 \right.
\nonumber \\
&\left.-2i\cl{D}_Ma_N[a^M,a^N]+\bar{\psi }\Gamma ^M[a_M,\psi ]
+2i\cl{D}_M\bar{c}\,[a^M,c]
-\frac{1}{2}\,[a^M,a^N]^2
 \right\}.
\end{align}
The first line defines the propagators of the fluctuation fields, while the second line represents various interaction vertices.

In our case the classical field  sits in the $\kkb$ corner of the $N\times N$ matrix (\ref{N+k-decomp}), the covariant derivative acts in the adjoint, and the $\kkb$ and $ \nnb$ components do not feel the background at all. They have the conventional $1/x^2$ propagators. The $\knb$ components do interact with the 't~Hooft line. Their propagators are of the form:
\begin{equation}
 \left\langle a_M^{1a}(x)a_N^{b1}(x')\right\rangle=\frac{\lambda \delta ^{ab}}{2N}\,G_{MN}(x,x'),
\end{equation}
where $a$ and $b$ are color indices and $G_{MN}$ is the Green's function in the monopole background. 

The Klein-Gordon operator in the field of the monopole:
\begin{equation}\label{D-squared}
 -\cl{D}^2=-\frac{\partial ^2}{\partial t^2}+H=\omega ^2+H,
\end{equation}
contains the Hamiltonian $H$:
\begin{equation}\label{DiracH}
 H=-\left(\vec{\partial }-i\vec{\cl{A}}\right)^2+\cl{\Phi }^2,
\end{equation}
where the $\cl{\Phi }^2$ term comes from the extra-dimensional part of the covariant derivative $D_I=-i\Phi _I^{\rm adj}$.
Once the eigenvalues and eigenfunctions of $H$ are known, the propagator can be reconstructed by spectral decomposition. 

The modes with precisely this kinetic operator will be called the ``easy" fields, as opposed to the ``hard" fields which in addition couple to $\cl{F}_{MN}$. The latter embodies interaction of spin with an external magnetic field, and is often called paramagnetic coupling. 

The background field strength has an obvious $\cl{F}_{ij}$ component, and vector fields do have paramagnetic interactions. But the R-symmetry dimensions are also magnetized, because of the non-zero scalar condensate (\ref{monopolePhi}). Indeed,
\begin{equation}
 \cl{F}_{iI}=\partial _i\cl{\Phi} _I=-q\,\frac{n_Ix_i}{r^3}\,,
\end{equation}
and the scalar mode proportional to $n_I$ ($\phi  _1$ in our notations) also has paramagnetic interactions through which it couples to the vector fields. We refer to paramagnetically coupled modes as hard fields. Finding their propagators boils down to solving a $4\times 4$ mixing problem. All in all we have:
\begin{align}
 &{\rm Easy:}\qquad a_0,\,c,\,\varphi _2,\ldots, \varphi_6,
\nonumber \\
&{\rm Hard:}\qquad a_i,\,\varphi _1.
\end{align}
We start with the simpler case of the easy fields.

\subsection{Remark on kinematics}

The 't~Hooft line preserves the $SO(2,1)\times SO(3)$ subgroup of conformal transformations. This is immediately visible in polar coordinates. The flat metric of  $\mathbbm{R}^4$ is conformally equivalent to that of $AdS_2\times S^2$:
\begin{equation}
 ds^2=r^2\left(\frac{dt^2+dr^2}{r^2}+d\theta ^2+\sin^2\theta \,d\varphi ^2\right).
\end{equation}
The prefactor is immaterial for conformal transformations and can be drop\-ped as far as symmetries are concerned,
whereupon $SO(2,1)\times SO(3)$ acts as the isometry group of  the metric. 

The emergent $AdS_2/CFT_1$ holography then defines an effective one-di\-men\-sional CFT on the 't~Hooft line.
The Kaluza-Klein reduction on $S^2$ (aka the partial wave decomposition) results in a field theory on $AdS_2$ with infinitely many fields whose masses map to the scaling dimensions by the usual holographic dictionary. As we shall see, for certain observables the KK expansion truncates and only the lowest partial wave survives. This phenomenon is to some extent similar to the Kondo effect \cite{Andrei:1980fv,vigman1980exact,wiegmann1981exact}, and we take the liberty to use this term once mode truncation is operative, a few examples of which will be discussed later in more detail.

By conformal symmetry the scalar propagator can only depend on the two cross-ratios:
\begin{align}
 \xi &=\frac{\left(t-t'\right)^2+r^2+r'{}^2}{2rr'},
\label{crossAdS2} \\
\eta& =\mathbf{n}\cdot \mathbf{n}',
\label{crossS2}
\end{align}
which are the geodesic distances on $AdS_2$ and $S^2$, respectively.

\subsection{Easy fields}

Quantum mechanics of a charged particle in the field of the Dirac monopole is a classic problem dating back to the work of Dirac  himself \cite{Dirac:1931kp}. Its solution is well-known \cite{Tamm:1931dda,Fierz:1944,Wu:1976qk} and can be pictured as a spectral flow of $SO(3)$ representations: the radial magnetic field of the monopole only affects the angular motion of the charged particle and, in effect, reshuffles different eigenstates of its† angular momentum  \cite{Wilczek:1981du}. The radial motion is affected  indirectly, through the centrifugal force.

The angular momentum in the presence of the monopole acquires an additional term \cite{Fierz:1944}:
\begin{equation}
 L_i=-i\varepsilon _{ijk}x_j\cl{D}_k-q\,\frac{x_i}{r}\,.
\end{equation}
It is this operator that commutes with the Hamiltonian (\ref{DiracH}), and so can be simultaneously diagonalized:
\begin{equation}
 \mathbf{L}^2Y_{\ell m}=\ell(\ell+1)Y_{\ell m},
 \qquad 
 L_zY_{\ell m}=mY_{\ell m}.
\end{equation}
The allowed values of angular momentum are \cite{Wu:1976qk}:
\begin{equation}\label{angularQ}
 \ell=q,q+1,q+2,\ldots ,
\end{equation}
which can be integer or half-integer, depending on the monopole charge.

The eigenfunctions $Y_{\ell m}$ are the monopole spherical harmonics \cite{Wu:1976qk,Wu:1977qk}, whose explicit form depends on the gauge. We adopt a slightly unconventional choice
\begin{equation}
 A=-q\cos\theta\, d\varphi ,
\end{equation}
with two symmetric Dirac strings. This is not minimal, one Dirac string can be gauged away, but our choice has an advantage of trivializing the $z$-component of the angular momentum: $L_z=-i\partial _\varphi $ and avoiding as a consequence some annoying phases in the Green's functions. 

The monopole harmonics are known explicitly and can be expressed through the Jacobi polynomials \cite{Tamm:1931dda}. With our choice of gauge, the unit-nor\-ma\-lized monopole harmonics are
\begin{equation}
 Y_{\ell m}(\theta ,\varphi )=\sqrt{\frac{(2\ell+1)(\ell-\alpha _+)!(\ell+\alpha _+)!}{4\pi (\ell+\alpha _-)!(\ell-\alpha _-)!}}\,
 \,{\rm e}\,^{im\varphi }\sin^\alpha \frac{\theta }{2}\,\cos^\beta \frac{\theta }{2}\,P_{\ell-\alpha _+}^{(\alpha ,\beta )}(\cos\theta ),
\end{equation}
where
\begin{equation}
 \alpha =|m+q|,\qquad \beta =|m-q|,\qquad \alpha _\pm=\frac{\alpha \pm \beta }{2}\,.
\end{equation}
The Jacobi polynomials are normalized such that $P_{n}^{(\alpha ,\beta )}(1)=1.$

This explicit form however is seldom needed. The single most important property of the monopole harmonics  is their addition theorem \cite{Wu:1977qk}:
\begin{equation}
 \sum_{m=-\ell}^{\ell}Y_{\ell m}(\mathbf{n})Y_{\ell m}^*(\mathbf{n}')
 =\frac{2\ell+1}{4\pi }\left(\frac{1+\mathbf{n}\cdot \mathbf{n}'}{2}\right)^q
 P_{\ell-q}^{(0,2q)}(\mathbf{n}\cdot \mathbf{n}'), \label{AdditiontheoremY}
\end{equation}
and the ensuing completeness condition:
\begin{equation}
 \sum_{m=-\ell}^{\ell}Y_{\ell m}(\mathbf{n})Y_{\ell m}^*(\mathbf{n})
 =\frac{2\ell+1}{4\pi }\,.
\end{equation}
The latter is the same for the usual spherical harmonics, while the addition theorem carries some dependence on the monopole charge.

Since
\begin{equation}\label{H-rad}
 H=\frac{1}{r}\,\left(-\frac{\partial ^2}{\partial r^2}+\frac{\mathbf{L}^2}{r^2}\right)r,
\end{equation}
partial wave decomposition separates variables in the Schr\"odinger equation\footnote{This form of the Hamiltonian, familiar as it is, is a result of a subtle cancellation: the centrifugal force  is actually proportional to $\mathbf{L}^2-q^2$ \cite{Fierz:1944,Wu:1976qk}, but the $q^2$ term exactly cancels agains an interaction with the scalar condensate, the last term in (\ref{DiracH}). This cancellation, a consequence of supersymmetry, is  the technical reason for many simplifications in the spectral problem.}. 
The radial wavefunctions are just the usual spherical Bessel functions, but with the angular momentum subject to novel quantization conditions  (\ref{angularQ}), the sole effect of the monopole background.

The complete set of delta-normalized eigenfunctions of $-\cl{D}^2$ is thus
\begin{equation}\label{easy-eigenmodes}
 \mathtt{eigenmodes}=\frac{1}{r}\,\,{\rm e}\,^{-i\omega t}\sqrt{kr}J_{j+\frac{1}{2}}(kr)Y_{jm}(\mathbf{n}),
\end{equation}
with the eigenvalues $\omega ^2+k^2$, and the angular momentum\footnote{We switched from $\ell$ to $j$ with a view towards fields with spin, for which the total and orbital angular momentum have to be distinguished. Furthermore,
we are here and in the following assuming that $q>0$.} taking values $j=q,q+1,\ldots $

The Green's function can be found by summing the eigenmodes:
\begin{eqnarray}\label{basicG}
  G(x,x')&=&\frac{1}{rr'}\,\sum_{jm}^{}Y^*_{jm}(\mathbf{n})Y_{jm}(\mathbf{n}')
\nonumber \\
&&\times 
\int_{-\infty }^{+\infty }\frac{d\omega}{2\pi }\, \int_{0}^{\infty }dk\,\,\frac{\,{\rm e}\,^{i\omega (t-t')}
  \sqrt{kr}\,J_{j+\frac{1}{2}}(kr)\sqrt{kr'}\,J_{j+\frac{1}{2}}(kr')
  }{\omega ^2+k^2}\,.
\end{eqnarray}
Taking into account that
\begin{equation}
 \int_{-\infty }^{+\infty }\frac{d\omega }{2\pi }\,\,\frac{\,{\rm e}\,^{-i\omega t}}{\omega ^2+k^2}=\frac{\,{\rm e}\,^{-k|t|}}{2k}\,,
\end{equation}
we find:
\begin{equation}\label{GfromD}
 G(x,x')=\frac{1}{rr'}\,\sum_{jm}^{}Y^*_{jm}(\mathbf{n})Y_{jm}(\mathbf{n}')D_{j+1}(|t-t'|,r,r'),
\end{equation}
where  
\begin{equation}\label{JJ-integral}
 D_\Delta  (t,r,r')=\frac{\sqrt{rr'}}{2}\int_{0}^{\infty }dk \,\,{\rm e}\,^{-kt}J_{\Delta -\frac{1}{2}} (kr)J_{\Delta -\frac{1}{2}} (kr').
\end{equation}

This function is the scalar propagator in $AdS_2$ for a field of mass
\begin{equation}
 m^2=\Delta (\Delta -1).
\end{equation}
By the usual AdS/CFT dictionary, $\Delta $ is the scaling dimension of the dual operator in ${\rm CFT}_1$. For the easy scalars a mode with angular momentum $j$ has $\Delta =j+1$. The spectrum thus starts with $\Delta =q+1$. For the smallest possible monopole charge the lowest dimension is equal to $3/2$. As we shall see the hard scalars actually contain modes with a lower dimension, $\Delta =1/2$, and for the fermions the lowest scaling dimension is $\Delta =1$.

The AdS propagator, nominally a function of three variables, only depends on the cross-ratio (\ref{crossAdS2}). Indeed, after performing the $k$-integral we get:
\begin{equation}\label{AdS2}
 D_\Delta  (|t-t'|,r,r')=\frac{1}{2\pi }\,Q_{\Delta -1} (\xi ),
\end{equation}
where $Q_\nu  $ is the Legendre function of the second kind.

Using the addition theorem for the monopole harmonics, the propagator can be written as
\begin{equation}\label{part-wave}
 G(x,x')=\frac{1}{4\pi rr'}\,\left(\frac{1+\eta }{2}\right)^q
 \sum_{j}^{}(2j+1)P_{j-q}^{(0,2q)}(\eta )D_{j+1}(\xi ).
\end{equation}
The $1/rr'$ pre-factor is prescribed by the conformal symmetry. The dependence on the two cross-ratios neatly factorizes.

\subsection{Dimensional regularization}

The preceding discussion applies to the four-dimensional 't~Hooft line. To use dimensional regularization we need propagators on $\mathbbm{R}^d$ with $d=4-2\varepsilon $. Dimensional regularization should preserve the quantized magnetic flux, and we define the $d$-dimensional counterpart of the 't~Hooft line as a $(d-3)$-dimensional monopole sheet. Co-dimension is then always three with the quantized magnetic flux piercing the $S^2$ that links the defect. The conformal equivalence
$$
 \mathbbm{R}^d\simeq AdS_{d-2}\times S^2
$$ 
carries over to an arbitrary dimension. We thus expect that the $AdS_2$ propagator (\ref{AdS2}) gets replaced by its $(d-2)$-dimensional counterpart:
\begin{equation}\label{Ddd}
 D_\Delta (\xi )=\frac{\Gamma (\Delta )}{2^{\Delta +1}\pi ^{\frac{d-3}{2}}\Gamma \left(\Delta -\frac{d-5}{2}\right)\xi ^\Delta }\,\,
 {}_2F_1\left(\frac{\Delta }{2}\,,\frac{\Delta +1}{2}\,;\Delta -\frac{d-5}{2};\frac{1}{\xi ^2}\right).
\end{equation}
The geodesic distance $\xi $ in $AdS_{d-2}$ is  the same as in (\ref{crossAdS2}) but with $t$ now understood as a $(d-3)$-dimensional vector. 

The regularized Green's function, to begin with, is given by
 (\ref{basicG}) with the frequency integration promoted to $(d-3)$ dimensions:
\begin{equation}
 \frac{d\omega }{2\pi }\rightarrow \frac{d^{d-3}\omega }{(2\pi )^{d-3}}\,.
\end{equation}
Repeating the steps leading to (\ref{GfromD}) we arrive at
\begin{equation}\label{spectral-dec-of-G}
 G(x,x')=\frac{1}{4\pi (rr')^{\frac{d}{2}-1}}
 \left(\frac{1+\eta }{2}\right)^q
 \sum_{j}^{}(2j+1)P_{j-q}^{(0,2q)}(\eta )D_{j+\frac{d}{2}-1}(\xi ),
\end{equation}
where
\begin{equation}\label{JJ-d}
 D_\Delta (t,r,r')=\left(\frac{rr'}{2\pi t}\right)^{\frac{d-3}{2}}t\int_{0}^{\infty }
 dk\,k^{\frac{d-3}{2}}K_{\frac{d-5}{2}}(kt)J_{\Delta -\frac{d-3}{2}}(kr)
 J_{\Delta -\frac{d-3}{2}}(kr').
\end{equation}
It can be checked that this function indeed integrates to (\ref{Ddd}).

It is important to emphasize that we have two sources of UV divergences here: short distance singularities from integration over $r$ and $t$, and divergent sums over angular momentum. Both are dimensionally regulated, but regularization cannot be removed until all integrals and all sums are calculated. 

\begin{figure}[t]
 \centerline{\includegraphics[width=2.5cm]{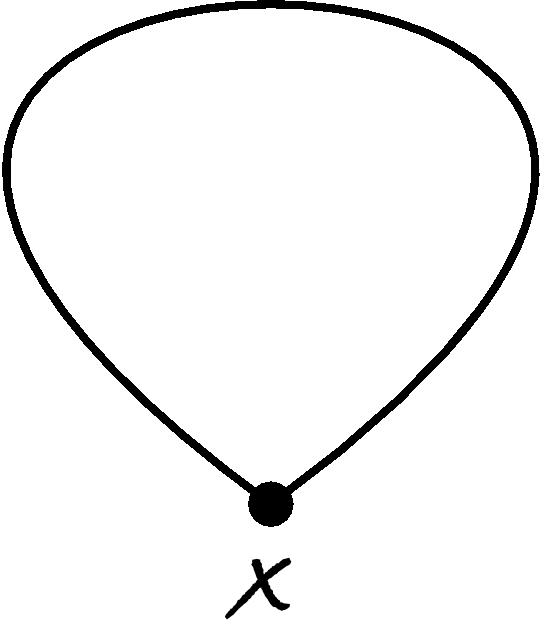}}
\caption{\label{Tadpole}\small The tadpole diagram.}
\end{figure}

To illustrate the point let us compute the simplest diagram, the tadpole (fig.~\ref{Tadpole}). This amounts to evaluating the propagator at coincident points $x'=x$. In terms of cross-ratios, we need to set $\xi =1=\eta $ which leads to log-divergences in the $AdS_2$ propagators. Indeed, the Green's function (\ref{JJ-integral}) is log-divergent when $t=0$, but the $AdS_{d-2}$ version (\ref{JJ-d}) is finite provided $d<4$. The  $\xi \rightarrow 1^-$  limit of (\ref{Ddd}) is then well-defined:

\begin{equation}
 D_\Delta (1)=\frac{\Gamma \left(2-\frac{d}{2}\right)\Gamma \left(\Delta \right)}{(4\pi )^{\frac{d}{2}-1}\Gamma \left(\Delta -d+4\right)}\,.
\end{equation}
The dimensionally-regularized tadpole thus equals to
\begin{equation}
{\tt tadpole}\equiv  G(x,x)=\frac{\Gamma \left(2-\frac{d}{2}\right)}{(4\pi )^{\frac{d}{2}}r^{d-2}}
 \sum_{j=q}^{\infty }(2j+1)\frac{\Gamma \left(j+\frac{d}{2}-1\right)}{\Gamma \left(j-\frac{d}{2}+3\right)}\,.
\end{equation}

There is an obvious $1/\varepsilon $ pole at $d\rightarrow 4$, but this is not the only source of infinities. Setting $d=4$ results in a divergent sum over partial waves  $\sum_{j}^{}(2j+1)$. It may be tempting to zeta-regularize the sum but that would lead to a wrong answer ($q+1/3$) for the residue of the $1/\varepsilon $ pole, not to say for the finite part.  The correct prescription is to compute the sum prior to taking the $d\rightarrow 4$ limit.

The sum is well-defined for $2-2q<d<2$ and can be analytically continued there from. It can be actually evaluated in the closed form:
\begin{equation}
 G(x,x)=-\frac{2q\Gamma \left(2-\frac{d}{2}\right)\Gamma \left(q+\frac{d}{2}-1\right)}{(4\pi )^{\frac{d}{2}}(d-2)\Gamma \left(q-\frac{d}{2}+2\right)r^{d-2}}.
\end{equation}
Now we can set $d=4-2\varepsilon $
and expand in $\varepsilon $:
\begin{equation}\label{G(x,x)}
 G(x,x)=-\frac{q^2}{16\pi ^2r^2}\left(\frac{1}{\varepsilon }+\ln(4\pi r^2)+1-\frac{1}{q}-2\psi (q)-\gamma \right).
\end{equation}
The residue is $q^2$, and vanishes for $q\rightarrow 0$, in accord with the absence of log-divergences in the background-free tadpole. The finite part also goes to zero at $q\rightarrow 0$ but at a slower rate.

\subsection{Hard fields}

The mixing between the quantum fluctuations of the hard scalar, $\phi_1$ and the spatial components of the
vector field, $\bf{a}$,  is encoded in the following $4\times 4$ matrix operator~\cite{Kristjansen:2023ysz}:
\begin{equation}\label{mixingM}
\widehat{M}
 \,=\, \frac{1}{r^2}
\begin{pmatrix}
r^2 \partial_t^2 +r^2 p_r^2+{\bf L}^2 & -i B \,{\bf n^T} \\
iB {\bf n}& r^2 \partial_t^2+ r^2 p_r^2+{\bf L}^2 -iB \, {\bf n \times} \end{pmatrix}.
\end{equation}
This operator is the counterpart of $\cl{D}^2$ from (\ref{D-squared}) with the paramagnetic coupling to $F_{MN}$ added, with radial and angular variables separated.
It acts on a column of type $(\phi_1,\bf{a})$ and can be inverted by spectral decomposition.

 To resolve the mixing we perform a Kaluza-Klein reduction
on $S^2$ introducing a mode expansion of the fields in terms of monopole spherical harmonics and monopole vector spherical harmonics respectively. Here, we shall work in the Weinberg basis of monopole vector spherical harmonics~\cite{Weinberg:1993sg} which 
most clearly reflects the $AdS_2\times S^2$ symmetry of the problem and furthermore aligns with the strategy for the treatment of fermions in a monopole potential~\cite{Kazama:1976fm} which we will follow in section~\ref{Fermions}.
An alternative basis, introduced in \cite{Olsen:1990jm}, of course gives the same result, see \cite{Kristjansen:2023ysz} for a detailed derivation.

The Weinberg vector monopole harmonics diagonalize  $\bf{J}^2$, $J_z$ and
$\bf{n}\cdot \bf{S}$:
\begin{eqnarray}
{\bf J}^2\, {\bf{C}}_{jm}^{(\lambda)}&=&j(j+1) \,{\bf C}_{jm}^{(\lambda)}, \\
 J_z\, {\bf{C}}_{jm}^{(\lambda)}&=& m \,{\bf C}_{jm}^{(\lambda)}, \\
 ({\bf n\cdot S})  \, {\bf{C}}_{jm}^{(\lambda)}&=&\lambda\,  {\bf{C}}_{jm}^{(\lambda)},
\end{eqnarray}
where $\lambda\in \{-1,0,1\}$. Notice that the Weinberg vector spherical harmonics do not diagonalize
the operator ${\bf L}^2$.

We thus expand our fields as follows
\begin{eqnarray}\label{hard-eigenmodes}
\phi_1(r,t,\theta,\phi)&=&e^{-i\omega t} \,\frac{1}{r} \,\sum_{jm}g_j(r) Y_{jm}(\theta,\phi), \label{phiexpansion} \\
{\bf a}(r, t, \theta,\phi)&=& e^{-i\omega t}\,\frac{1}{r} \,\sum_{jm} \left(f^{(-1)}_j(r) \,{\bf C}_{jm}^{(-1)}+f^{(0)}_j(r)\, {\bf C}_{jm}^{(0)}+
f_j^{(+1)}(r) \, {\bf C}_{jm}^{(+1)}\right), \nonumber  \\
\label{aexpansion}
\end{eqnarray}
where the $Y_{jm}$ are the scalar monopole spherical harmonics discussed above. We wish to determine
$g_j(r)$, $f_j^{(-1)}(r)$, $f_j^{(0)}(r)$, $f_j^{(+1)}(r)$ such that the configuration $(\phi_1,{\bf a})$ is  an eigenmode of the matrix operator (\ref{mixingM}) with the eigenvalue $k^2+\omega^2$. 
To this end, we use the decomposition in the spherical harmonics along with
the following relations:
\begin{eqnarray}
{\bf n}\cdot {\bf{C}}_{jm}^{(\lambda)}&=&\delta_{\lambda 0} Y_{jm}, \\
{\bf n} \times {\bf{C}}_{jm}^{(\lambda)} &=& -i\lambda \,{\bf{C}}_{jm}^{(\lambda)},
\end{eqnarray}
reflecting the fact that ${\bf{C}}_{jm}^{(0)}$ is purely radial while ${\bf{C}}_{jm}^{(\pm 1)}$ are transverse. In fact,
\begin{equation}
{\bf C}_{jm}^{(0)}={\bf n} Y_{jm}.
\end{equation}

As demonstrated in~\cite{Kristjansen:2023ysz}, it is consistent to assume that all four unknown functions above can be expressed in terms of the same Bessel function, more precisely that
\begin{equation}\label{Bessel-ansatz}
g_j(r)= g_j \, \sqrt{kr}\, J_{\nu}(k r), \hspace{0.5cm}
f_j^{(\lambda)}(r)= f_j^{(\lambda)}\sqrt{kr} \, J_{\nu}(k r),
\end{equation}
where $g_j$ and  $f_j^{(\lambda)}$  are just constants. For an easy scalar the index $\nu $ equals $j+1/2$, cf.~(\ref{easy-eigenmodes}). Here, as a result of mixing, the index will take four different values for a given angular momentum:      
\begin{equation}
 \nu =j+\frac{1}{2}+\delta _\alpha ,\qquad \alpha =1,\ldots ,4.
\end{equation}

The possible
values of $\nu$ for a given value of $j$ are determined by the condition that the quantity 
\begin{equation}
a=\nu^2-\left(j+\frac{1}{2}\right)^2,
\end{equation}
must be an eigenvalue of the following matrix 
\begin{equation}
\begin{pmatrix} \nonumber
0 & 0 & 2 i q & 0 \\
0 & 0 & \sqrt{2(j-q)(1+j+q)} &0 \\
-2i q &  \sqrt{2(j-q)(1+j+q)} & -2 & \sqrt{2(1+j-q)(j+q)} \\
0 & 0 & \sqrt{2(1+j-q)(j+q)} &0 
\end{pmatrix}, 
\end{equation}
and the  associated eigenvector then consists of the corresponding expansion coefficients $(g_j, f_j^{(-1)}, f_j^{(0)}, f_j^{(+1)})$. This follows by acting of the operator (\ref{mixingM}) on (\ref{phiexpansion}) and~(\ref{aexpansion}), taking the Bessel-function ansatz (\ref{Bessel-ansatz}) and requiring the result to be an eigenfunction with the eigenvalue $k^2+\omega^2$. The explicit form of $\mathbf{L}^2$ necessary for this calculation can be found in \cite{Weinberg:1993sg}.

The above condition results in the following set of possible values for $\nu$
\begin{equation}
\nu=\left\{j-\frac{1}{2}, j+\frac{1}{2}, j+ \frac{1}{2}, j+\frac{3}{2}\right\}, 
\end{equation} 
or, in other words,
\begin{equation}
 \delta _\alpha =\left\{-1,0,0,1\right\}.
\end{equation}
It is quite remarkable that the indices $\delta _\alpha $ come out as simple integers independent of $j$. Vector modes of a non-supersymmetric monopole have a considerably more complicated spectrum  \cite{Olsen:1990jm}. In that case mixing results in a quadratic equation for $\nu $ with explicit radicals in the solution  \cite{Olsen:1990jm}. Coupling to the scalar adds an extra mode making the mixing problem more complicated but results in a simpler spectrum. This simplicity, we believe, can be traced back to supersymmetry preserved by the monopole background.

The normalized eigenvectors corresponding to the eigenvalues above read
\begin{equation}\label{vect12}
\begingroup
\renewcommand*{\arraystretch}{1.7}
v_1=\begin{bmatrix}
-i\sqrt{\frac{(j^2-q^2)}{j(1+2j)}} \\ 
\sqrt{\frac{(j+q)(1+j+q)}{2j(1+2j)}} \\
0\\  -\sqrt{\frac{(j-q)(1+j-q)}{2j(1+2j)}}
\end{bmatrix}, \hspace{0.5cm}
v_2= \begin{bmatrix}
i \sqrt{\frac{(1+j-q)(1+j+q)}{(1+j)(1+2j)}} \\
  \sqrt{\frac{(j-q)(1+j-q)}{2(1+j)(1+2j)}} \\
0 \\
 -\sqrt{\frac{(j+q)(1+j+q)}{2(1+j)(1+2j)}}
\end{bmatrix}, 
\endgroup
\hspace{0.5cm} \delta_{1,2} =0,
\end{equation}

\begin{equation}\label{vect34}
\begingroup
\renewcommand*{\arraystretch}{1.7}
v_3=\begin{bmatrix}
\frac{iq}{\sqrt{j(1+2j)}} \\
\sqrt{\frac{j^2+j-q^2-q}{j(1+2j)}} \\
\sqrt{\frac{j}{1+2j}} \\
\sqrt{\frac{j^2+j-q^2+q}{2j(1+2j)}}
\end{bmatrix}, \hspace{0.5cm} \delta_3 =-1, \hspace{0.5cm}
\renewcommand*{\arraystretch}{1.7}
v_4= \begin{bmatrix}
\frac{iq}{\sqrt{(1+j)(1+2j)} }\\
\sqrt{\frac{j+j^2-q^2-q}{2(1+j)(1+2j))}} \\
-\sqrt{\frac{1+j}{1+2j}} \\
\sqrt{\frac{j^2+j-q^2+q}{2(1+j)(1+2j)}}
\end{bmatrix},
 \hspace{0.5cm} \delta_4 =1.
\endgroup
\end{equation}
These expressions are valid for generic $j$ and $q$. There are two special cases, $j=q>0$ and $j=q-1\geq 0$. For $j=q$, the  Weinberg vector harmonic with $\lambda=-1$ does not exist, and indeed the second row and the second column of the mixing matrix 
vanish in this case. The eigenvectors $v_1$ and $v_2$ get replaced by a single vector, $v$, whereas the eigenvectors $v_3$
and $v_4$ attain their limiting values. More precisely, for $j=q$, the eigenvectors read
\begin{equation}
v=\begin{bmatrix}
-\frac{i}{\sqrt{1+j} }\\
0 \\
\frac{\sqrt{j}}{\sqrt{(j+1)}}
\end{bmatrix}, \hspace{0.5cm}\delta =0,
\end{equation}
and 
\begin{equation}
\begingroup
\renewcommand*{\arraystretch}{1.7}
v_3=\begin{bmatrix}
\frac{i\,\sqrt{j}}{\sqrt{1+2j} }\\
\frac{\sqrt{j}}{\sqrt{1+2j}} \\
\frac{1}{\sqrt{1+2j}}
\end{bmatrix}, \hspace{0.5cm}\delta _3=-1, \hspace{0.5cm}
v_4=\begin{bmatrix}
\frac{i\,j}{\sqrt{(1+j)(1+2j)} }\\
-\frac{\sqrt{1+j}}{\sqrt{1+2j}} \\
\frac{\sqrt{j}}{\sqrt{(1+j)(1+2j)}}
\end{bmatrix}, \hspace{0.5cm}\delta _4=1.
\endgroup
\end{equation}

For $j=q-1$ the scalar vector harmonic does not exist and the Weinberg vector harmonic has only one component
${\bf C}_{q-1m}^{(+1)}$. In this case one finds
\begin{equation}
f_j^{{(+1)}}(r)=\sqrt{kr}\, J_{j+\frac{1}{2}}(kr).
\end{equation}

\subsection{Kondo effect}

Kondo effect refers to scattering of conduction electrons off a magnetic impurity, with only the s-wave scattering relevant in the RG sense.  The problem effectively becomes $1+1$ dimensional and eventually solvable by Bethe ansatz \cite{Andrei:1980fv,vigman1980exact,wiegmann1981exact}. Scattering of a charge on a monopole in the $\mathcal{N}=4$ SYM has some loose parallels to this phenomenon. There is no RG running in this case, but supersymmetry cancellations lead to only the lowest angular harmonics contributing to some select observables. A few examples of such dynamical reduction to 1+1 dimensions are given below and by certain abuse of terminology we also call them Kondo effect.

\subsubsection{Propagator of hard scalar}

Due to mixing with vectors the hard scalar has four components with the $AdS_2$ scaling dimensions $\Delta _{\alpha} =j+1+\delta _\alpha $, $\delta _\alpha $ taking values $\left\{-1,0,0,1\right\}$. Each component enters with the weight 
$|v_\alpha ^1|^2$ given by (\ref{vect12}), (\ref{vect34}), and $j$ runs over the range $q, q+1,\ldots $
The lowest scaling dimension of a scalar operator thus equals the monopole charge $q$, in the minimal case the smallest scaling dimension is $\Delta =1/2$.

The dimensionally regularized hard-scalar propagator takes the following form:
\begin{equation}
 \widetilde{G}(x,x')=\frac{1}{4\pi (rr')^{\frac{d}{2}-1}}
 \left(\frac{1+\eta }{2}\right)^q
 \sum_{j }^{}(2j+1)P_{j-q}^{(0,2q)}(\eta )\sum_{\alpha }^{}|v_\alpha ^1|^2D_{j+\frac{d}{2}-1+\delta _\alpha }(\xi ).
\end{equation}
Reading off the weights from (\ref{vect12}), (\ref{vect34}), we get:
\begin{align}\label{hard-spectral}
 \widetilde{G}(x,x')=&\frac{1}{4\pi (rr')^{\frac{d}{2}-1}}
 \left(\frac{1+\eta }{2}\right)^q
 \sum_{j }^{}P_{j-q}^{(0,2q)}(\eta )
\nonumber \\ 
 &\times 
 \left[
 \frac{q^2}{j}\, D_{j+\frac{d}{2}-2 }
 +
 \frac{(2j+1)(j^2+j-q^2)}{j(j+1)}\,D_{j+\frac{d}{2}-1 }
 +
  \frac{q^2}{j+1}\, D_{j+\frac{d}{2} }
 \right].
\end{align}

\subsubsection{Anomalous propagator}\label{anG-sec}

Absent the monopole, the easy and hard propagators are no different from the conventional Klein-Gordon Green's function\footnote{This may not be immediately obvious from expressions like (\ref{part-wave}) and (\ref{hard-spectral}). Just for completeness, we show  in  appendix~\ref{part-wave-app} how the standard propagator is recovered from the partial wave decomposition.} $1/4\pi ^2|x-x'|^2$. Their difference thus measures genuine correlations induced by the 't~Hooft loop. We thus define the anomalous propagator
\begin{equation}\label{anG}
 G^-=\widetilde{G}-G,
\end{equation}
which vanishes at $q=0$ when the external field is switched off.
This function will feature prominently in later calculations and deserves a more in-depth discussion.

From (\ref{spectral-dec-of-G}), (\ref{hard-spectral}) we find:
\begin{align}
 G^-(x,x')=&\frac{q^2}{4\pi (rr')^{\frac{d}{2}-1}}
 \left(\frac{1+\eta }{2}\right)^q
 \nonumber \\
 &\times 
 \sum_{j }^{}P_{j-q}^{(0,2q)}(\eta )
 \left(
  \frac{D_{j+\frac{d}{2}-2}-D_{j+\frac{d}{2}-1}}{j}
  -\frac{D_{j+\frac{d}{2}-1}-D_{j+\frac{d}{2}}}{j+1}
 \right).
\end{align}
This vanishes at $q=0$ as expected, and simplifies quite a lot compared to the hard or easy propagators alone.

The most drastic simplification occurs in the collinear kinematics when the endpoints of the propagator are coplanar with the 't~Hooft line, then $\eta =1$ and $P_{j-q}^{(0,2q)}=1$. The partial-wave sum then telescopes leaving behind the lowest-harmonics contribution with $j=q$:
\begin{equation}\label{tildeGeta=1}
 \left.\vphantom{\frac{1}{2}}G^-(x,x')\right|_{\eta =1}=\frac{q}{4\pi (rr')^{\frac{d}{2}-1}}
 \left(
  D_{q+\frac{d}{2}-2}-D_{q+\frac{d}{2}-1}
 \right).
\end{equation}
This is the Kondo effect -- only the lowest mode  survives, and we get only one field in $AdS_2$ instead of a whole  KK tower.

We do not know the precise reason for this cancellation, but suspect supersymmetry as an undelying reason.  It is also unclear to us why the cancellation is incomplete and the lowest harmonics gives a non-zero contribution. Presumably this has to do with the multiplet structure of the $\mathfrak{osp}(4^*|4)$ supersymmerty, the lowest partial wave satisfying some sort of shortening condition.

The coincidence limit of the anomalous propagator is non-singular and we can drop dimensional regularization altogether:
\begin{equation}\label{anom-Green}
 G^-(x,x)=\frac{2q\Gamma \left(3-\frac{d}{2}\right)\Gamma \left(q+\frac{d}{2}-2\right)}{(4\pi )^\frac{d}{2}\Gamma \left(q-\frac{d}{2}+3\right)r^{d-2}}
 \stackrel{d\rightarrow 4}{= }\frac{1}{8\pi ^2r^2}\,.
\end{equation}
Notice however that the limits $x'\rightarrow x$, $d\rightarrow 4$ and $q\rightarrow 0$ do not commute.

\subsection{Fermions \label{Fermions}}

We now turn to fermions. Angular momentum quantization on the monopole background allows for the s-wave scattering for half-integer spin. An enhanced amplitude of this process
underlies the famous Callan-Rubakov effect \cite{Rubakov:1981rg,Callan:1982ah,Rubakov:1982fp,Callan:1982au,Callan:1982ac}
whose puzzling features \cite{Kazama:1976fm} are
 continued to be discussed to this day \cite{Brennan:2021ewu,vanBeest:2023dbu,Khoze:2023kiu,Brennan:2023tae,vanBeest:2023mbs}. In the language of dCFT, the Callan-Rubakov effect is caused by a  fermion operator localized on the defect of exact dimension $\Delta =1/2$ \cite{Aharony:2023amq,vanBeest:2023dbu}, appropriate for a free $1+1$d fermion, hence a propagating fermion zero mode. The supersymmetry-breaking 't~Hooft line in the $\mathcal{N}=4$ SYM, studied in \cite{Aharony:2023amq}, indeed carries fermion zero modes. 
 
 As we shall see, the Callan-Rubakov effect does not occur for the supersymmetric 't~Hooft loops, because the scalar condensate (\ref{monopolePhi}) repels fermions from the monopole core. The underlying reason is of course supersymmetry, a fermion operator of dimension $\Delta =1/2$ would require a matching bosonic counterpart, which is simply not there, and is probably inconsistent with the stability of the 't~Hooft line. We will find that wavefunctions of all fermion modes vanish at the monopole core, the lowest fermion operator having dimension $\Delta =1$  (for $q=1/2$).

The quadratic terms for the fermions read
\begin{equation}
S^{ferm}_2=\frac{1}{2}\mbox{tr}\left(
i\bar{\Psi}\Gamma^\mu \partial_\mu \Psi+\bar{\Psi} \Gamma^1\left[\cl{\Phi}_1,\Psi\right]+\bar{\Psi}\Gamma^\mu \left[\cl{A}_\mu,\Psi\right]
\right),\label{S2ferm}
\end{equation}
where the $\{\Gamma^\mu,\Gamma^i\}$ are ten-dimensional $\Gamma$-matrices with $\mu=0,1,2,3$ and 
 $i=1,2,\ldots, 6$.
We  start from the following  representation of the  four-dimensional $\gamma$-matrices
\begin{equation}
\gamma^0=
\begin{pmatrix}
I & 0\\
0 & -I
\end{pmatrix}, \hspace{0.5cm}
\gamma^j=
\begin{pmatrix}
 0 &-i \sigma_j\\
i\sigma_j & 0
\end{pmatrix}, \hspace{0.5cm} \gamma^5=
\begin{pmatrix}
 0 & I\\
I & 0
\end{pmatrix},
\end{equation}
and define the ten-dimensional gamma matrices as in eqn.~(\ref{10D-Dirac}).
The matrix $\hat{\gamma}^1$ squares to one and has eigenvalues $\pm 1$ with an equal number of each type. We can thus start by
diagonalizing $\hat{\gamma}^1$ and consider the eigenspaces corresponding the the eigenvalues $+1$ and $-1$ separately.
Reducing to four dimensions we are left with four four-component Majo\-ra\-na-Weyl spinors with $N\times N$ matrix indices.
We notice that the classical fields only have a non-vanishing 11-component and thus couple only to field components 
$\Psi_{1j}$ and $\Psi_{j1}$ for $j\neq 1$. Hence, we have $(N-1)$ contributions to the action with non-trivial mass  terms of the form
\begin{equation}
S_2^f ={\Psi}^ \top C \left(i \gamma^\mu  \left( \partial_{\mu}-i\cl{A}_{\mu}\right) \pm \gamma^5 \,\frac{q}{r}\right) \Psi,
\label{S2f}
\end{equation}
where the $\pm$ corresponds to the two eigenspaces of $\hat{\gamma}^1$.
Introducing the explicit representation of the four-dimensional gamma matrices and considering to begin with the case
of the positive sign of the off-diagonal elements 
the action~(\ref{S2f}) is expressed as
\begin{equation}
S_2^{f}=
\begin{pmatrix}
{\Psi}_A ^\top,{\Psi}_B ^\top 
\end{pmatrix} C
\begin{pmatrix}
i\partial_t &  \vec{\sigma}\cdot\left(\vec{\partial}-i\vec{\cl{A}}\right)+ \frac{q}{r}\\
 -\vec{\sigma}\cdot\left(\vec{\partial}-i\vec{\cl{A}}\right) + \frac{q}{r} & -i\partial_t
\end{pmatrix}
\begin{pmatrix}
\Psi_A \\
\Psi_B
\end{pmatrix}. \nonumber
\end{equation}
In order to diagonalize $S_2^f$ we will make use of the  spinor monopole harmonics, developed in~\cite{Kazama:1976fm}.  Placing a particle in the monopole potential shifts
its angular momentum by $q$. \label{spinorharmonics}
For each value of total angular momentum $j$ there are two independent spinor monopole harmonics that are constructed by combining the usual scalar spherical harmonics with the spin-$1/2$ representation. They take the form~\cite{Kazama:1976fm}
\begin{equation}
\begingroup
\renewcommand*{\arraystretch}{1.7}
\phi_{jm}^{(1)}=\begin{bmatrix}
\left(\frac{j+m}{2j}\right)^{1/2} Y_{j-1/2\,m-1/2} \\
\left(\frac{j-m}{2j}\right)^{1/2} Y_{j-1/2\,m+1/2}
\end{bmatrix}, \hspace{0.5cm}
\phi_{jm}^{(2)}=
\begin{bmatrix}
-\left(\frac{j-m+1}{2j+2}\right)^{1/2} Y_{j+1/2\,m-1/2}\\
\left(\frac{j+m+1}{2j+2}\right)^{1/2} Y_{j+1/2\,m+1/2}
\end{bmatrix},
\endgroup
\end{equation}
and are eigenstates of ${\bf J}^2$, $J_z$ and ${\bf L}^2$, in particular
\begin{equation}
{\bf L}^2\phi_{jm}^{(1)}= \left(j-\frac{1}{2}\right)\left(j+\frac{1}{2}\right)\phi_{jm}^{(1)}, \hspace{0.5cm}\nonumber
{\bf L}^2\phi_{jm}^{(2)}= \left(j+\frac{1}{2}\right)\left(j+\frac{3}{2}\right)\phi_{jm}^{(2)}, \nonumber
\end{equation}
 For the special
case $j=q-\frac{1}{2}\geq 0$, only $\phi_{jm}^{(2)}$ exists. 

As in the case of the vector monopole spherical harmonics it is possible to change to a basis of spinor spherical harmonics 
which are eigenstates of ${\bf {n}\cdot S }$ instead of ${\bf L}^2$ and thus reflect more directly the $AdS_2\times S^2$
symmetry of the problem.  These spinors, defined for $j>q-\frac{1}{2}$,  fulfil the relations
\begin{equation}
{\bf  \bf{n}\cdot S }\, \chi_{jm}^{(1)}=- \chi_{jm}^{(1)}, \hspace{0.5cm}
{\bf  \bf{n}\cdot S }\, \chi_{jm}^{(2)}=\chi_{jm}^{(2)},
\end{equation}
and are related to the spinor harmonics $\phi_{jm}^{(1)}$, $\phi_{jm}^{(2)}$ in the following way
\begin{eqnarray}
\chi_{jm}^{(1)}&=&c\,\phi_{jm}^{(1)}+s \,\phi_{jm}^{(2)}, \\
\chi_{jm}^{(2)}&=&-s\,\phi_{jm}^{(1)}+c \,\phi_{jm}^{(2)}.
\end{eqnarray}
Both the $\phi$'s and the $\chi$'s constitute a complete orthonormal set, and the constants $c$ and $s$ can be viewed as 
$\cos\varphi_0$ and $\sin\varphi_0$ for some angle $\varphi_0$. They are given by 
\begin{eqnarray}
c&\equiv&\cos \varphi_0=\frac{1}{\sqrt{2}}\frac{(j+\frac{1}{2}+q)^{1/2}}{(j+\frac{1}{2})^{1/2}}, \label{cosinus}
\\
s&\equiv& \sin\varphi_0= \frac{1}{\sqrt{2}}\frac{(j+\frac{1}{2}-q)^{1/2}}{(j+\frac{1}{2})^{1/2}}. \label{sinus}
\end{eqnarray}
For $j=q-\frac{1}{2}$ one continues to work with only $\phi_{jm}^{(2)}\equiv \chi_{m}$ which fulfills
\begin{equation}
{\bf  {n}\cdot S }\, \chi_{m}=\chi_{m}. \label{lowestmode}
\end{equation}
 The $\chi$-functions obey the following relations
\begin{eqnarray}
\vec{\sigma}\cdot\left(\vec{\partial}-i\vec{\cl{A}}\right)\,f(r)\chi_{jm}^{(1)}&=
& -\left(\partial_r+ r^{-1}\right) f(r)\, \chi_{jm}^{(1)}+\mu r^{-1} f(r)\, \chi_{jm}^{(2)},
\\
\vec{\sigma}\cdot\left(\vec{\partial}-i\vec{\cl{A}}\right)\,g(r)\chi_{jm}^{(2)}&=& 
+\left(\partial_r+ r^{-1}\right)\, g(r) \,\chi_{jm}^{(2)}-\mu r^{-1} g(r)\, \chi_{jm}^{(1)},
\end{eqnarray}
where $f(r)$ and $g(r)$ are arbitrary functions of $r$ and 
\begin{equation}
\mu=\left(\left(j+\frac{1}{2}\right)^2-q^2\right)^{1/2}.
\end{equation}
The similar equation for $\chi_m$ reads
\begin{equation}
\vec{\sigma}\cdot\left(\vec{\partial}-i\vec{\cl{A}}\right)\,f(r)\,\chi_{m}= (\partial_r+r^{-1}) f(r)\, \chi_m,
\end{equation}
again with $f(r)$ an arbitrary function of $r$. 

After Fourier transform in $t$, we need to solve the eigenvalue problem
\begin{equation}\label{eigenvalueproblem}
\begin{pmatrix}
E-\lambda & \vec{\sigma}\cdot\left(\vec{\partial}-i\vec{\cl{A}}\right)+ \frac{q}{r} \\
-\vec{\sigma}\cdot\left(\vec{\partial}-i\vec{\cl{A}}\right) + \frac{q}{r}& -E-\lambda
\end{pmatrix}
\begin{pmatrix}
\Psi_A \\
\Psi_B
\end{pmatrix}=
\begin{pmatrix}
0 \\
0
\end{pmatrix}.
\end{equation}
In the following we will restrict ourselves to considering  eqn.~(\ref{eigenvalueproblem}) with the plus sign. First, we consider the
generic case $j> q-\frac{1}{2}$, and we make the 
 ansatz 
\begin{equation}
\begingroup
\renewcommand*{\arraystretch}{1.5}
\begin{pmatrix}
\Psi_A \\
\Psi_B
\end{pmatrix}= 
\begin{pmatrix}
f_-(r) \chi^{(1)}_{jm}+f_+(r) \chi^{(2)}_{jm} \\
g_-(r) \chi^{(1)}_{jm}+g_+(r) \chi^{(2)}_{jm}
\end{pmatrix}.
\endgroup
\end{equation}
Writing out the four equations which result from eqn.~(\ref{eigenvalueproblem}) we find
\begin{eqnarray}
0&=& (E-\lambda)f_--\mu r^{-1} g_+-(\partial_r+r^{-1}) g_-+\frac{q}{r}\,g_-,\label{eqnfg1} \\
0&=&  (E-\lambda)f_++\mu r^{-1} g_-+(\partial_r+r^{-1}) g_++\frac{q}{r}\,g_+,\label{eqnfg2}\\
0&=& - (E+\lambda)g_-+\mu r^{-1} f_++(\partial_r+r^{-1}) f_-+\frac{q}{r}\,f_-,\label{eqnfg3}\\
0&=& -(E+\lambda)g_+-\mu r^{-1} f_--(\partial_r+r^{-1}) f_++\frac{q}{r}\,f_+.\label{eqnfg4} 
\end{eqnarray}
 We expect that the result should be expressible in terms of the two 
Bessel functions $J_{\nu_+}$ and $J_{\nu_-}$ where
\begin{equation}
\nu_+=j+1, \hspace{0.5cm} \nu_-=j.
\end{equation}
Therefore, we make the further specification 
\begin{eqnarray}
f_+(r)&=& f_{++}\, {\cal J}_+(r)
+f_{+-}\,{\cal J}_-(r), \label{spec1}\\
f_-(r)&=& f_{-+}\, {\cal J}_+(r)
+f_{--}\,{\cal J}_-(r), \label{spec2}\\
g_+(r)&=& g_{++}\, {\cal J}_+(r)
+g_{+-}\,{\cal J}_-(r), \label{spec3}\\
g_-(r)&=& g_{-+}\, {\cal J}_+(r)
+g_{--}\,{\cal J}_-(r), \label{spec4}
\end{eqnarray}
where
\begin{equation}\label{calJ}
{\cal J}_{\pm}(r)= \frac{1}{\sqrt{kr}}\, J_{\nu_\pm}(kr), \hspace{0.5cm} k^2=\lambda^2-E^2,
\end{equation}
and where $f_{++}, f_{+-}$ etc. are constants. 
We now notice that
\begin{eqnarray}
\partial_r \,{\cal J}_{+}(r)&=&-\left(J+\frac{3}{2}\right)r^{-1} {\cal J}_{+}(r)+k {\cal J}_{-}(r), \\
\partial_r \,{\cal J}_{-}(r)&=&\left(J-\frac{1}{2}\right)r^{-1} {\cal J}_{-}(r)-k {\cal J}_{+}(r),
\end{eqnarray}
and inserting the more specific ans\"{a}tze  from equations (\ref{spec1})--(\ref{spec4}) 
into eqns.\ (\ref{eqnfg1})-(\ref{eqnfg4})
we get the
following relations between the various coefficients
\begin{eqnarray}
f_{+-}=\frac{\lambda+E}{k}\,\,g_{++},&& \hspace{0.5cm} f_{++}=-\frac{\lambda+E}{k}
\,\, g_{+-},\\
\hspace{0.5cm}f_{-+}=\frac{\lambda+E}{k}
\,\,g_{--},&&\hspace{0.5cm}f_{--}=-
\frac{\lambda+E}{k}\,\,g_{-+},
\end{eqnarray}
\begin{equation}
g_{-+}=t\,g_{++},\hspace{0.5cm} g_{+-}=-t\,g_{--},\hspace{0.5cm}
\end{equation}
where
\begin{equation}
t\equiv \tan\varphi_0= \sqrt{\frac{j+\frac{1}{2}-q}{j+\frac{1}{2}+q}}. \label{tangent}
\end{equation}
Making use of these identities we can now write our general solution as
\begin{equation}
\begingroup
\renewcommand*{\arraystretch}{1.3}
\begin{pmatrix}
\Psi_A \\
\Psi_B
\end{pmatrix} =
e^{iEt}
\begin{pmatrix}
\left.\left.
\frac{\lambda+E}{k}
 \, \right\{ g_{--}
{\cal J}_{+}(r)\left[c\,\chi^{(1)}+s\,\chi^{(2)}\right]+g_{++}{\cal J}_-(r)
 \left[- s\, \chi^{(1)}+c\,\chi^{(2)}\right] \right\} \\
g_{++}{\cal J}_{+}(r)\left[s\,\chi^{(1)}+c\,\chi^{(2)}\right]+g_{--}{\cal J}_-(r)
 \left[ c\, \chi^{(1)}-s\,\chi^{(2)}\right]
\end{pmatrix}, \nonumber \\
\endgroup
\end{equation}
which we can also express as
\begin{equation}
\begingroup
\renewcommand*{\arraystretch}{1.3}
\begin{pmatrix}
\Psi_A \\
\Psi_B
\end{pmatrix} =g_{--} \,
\begin{pmatrix}
\frac{\lambda+E}{k}\,{\cal J}_{+}(r)\chi_+^1
  \\
{\cal J}_-(r)\chi_-^1
\end{pmatrix}e^{iEt}+
g_{++}\begin{pmatrix}
\frac{\lambda+E}{k}\,{\cal J}_-(r) \chi_+^2
  \\
{\cal J}_{+}(r)\chi_-^2
\end{pmatrix}e^{iEt},
\nonumber
\endgroup
\end{equation}
where we notice that $(\chi_-^1,\chi_-^2)$ and  $(\chi_+^1,\chi_+^2)$ appear from 
 $(\chi^1,\chi^2)$ by a rotation with angle $-\varphi_0$ and $+\varphi_0$ respectively. The integration constants should
 be fixed by normalization $\int\Psi^{\dagger} \Psi\sim 1$ and we choose to work with the following set of 
 eigenfunctions corresponding to the
 eigenvalue $\lambda$
\begin{equation}
\Psi_1=\left(\frac{k}{2\lambda}\right)^{1/2}
\begin{pmatrix}
  \sqrt{\lambda+E}\,{\cal J}_{+}(r)\chi_+^1
  \\
  \sqrt{\lambda-E}
\,{\cal J}_-(r)\chi_-^1
\end{pmatrix}e^{iEt}, \hspace{0.5cm}
\Psi_{2}=\left(\frac{k}{2\lambda}\right)^{1/2}\begin{pmatrix}
 \sqrt{\lambda+E}\,{\cal J}_-(r) \chi_+^2
  \\
 \sqrt{\lambda-E}
\,{\cal J}_{+}(r)\chi_-^2
\end{pmatrix}e^{iEt}.
\end{equation}
We stress that the Bessel functions which appear in the expressions above have the integer indices $j+1$ and $j$. This
is in contrast to the case of a fermion in a standard monopole potential with no coupling to a scalar where the Bessel
functions which appear have non-integer and in general irrational indices~\cite{Kazama:1976fm}.  A similar effect was seen for the vector particle and is a consequence of the supersymmetry of ${\cal N}=4$ SYM.  
It is straightforward to find the 
corresponding wave functions for the case where the off-diagonal elements $\frac{q}{r}$ in eqn.~(\ref{eigenvalueproblem}) have the opposite sign
as  this only leads to a change of the sign of $q$ in eqn.~(\ref{tangent}).

At the origin, the fermion field behaves as 
$$\Psi \sim \frac{1}{r^{3/2}}\,r^{\nu _\pm+1}.$$
The bulk-to-boundary OPE then implies the existence of defect operators of dimension $\Delta =\nu _\pm+1=j+2$ and $j+1$. We remind that here $j=q+1/2, q+3/2,\ldots $

We still need to  find the appropriate contribution for the lowest fermionic mode
where there is only one independent spinor spherical harmonic. In that case we make the following assumption about
the wave function
\begin{equation}
\begin{pmatrix}
\Psi_A \\
\Psi_B
\end{pmatrix}= 
\begin{pmatrix}
\left[f_+ \,{\cal J_+}(r)+f_- \,{\cal J}_-(r)\right] \,\chi_{m} \\
\left[g_+\, {\cal J_+}(r)+g_- \,{\cal J}_- (r)\right]\, \chi_{m}
\end{pmatrix} e^{iEt},
\end{equation}
where ${\cal J}_\pm$ were given in~(\ref{calJ}) with $j=q-\frac{1}{2}$.
  Plugging the ansatz for the wave function
into the eigenvalue equation~(\ref{eigenvalueproblem}), we find that the form of the solution depends on the sign of the off-diagonal
terms of type $\pm \frac{q}{r}$. In the case of the plus sign we arrive at the following expression for the normalized
wave function
\begin{equation}
\Psi_+= 
\left(\frac{k}{2\lambda}\right)^{1/2}\begin{pmatrix}
\sqrt{\lambda+E}\,{\cal J_-}(r) \,\chi_{m} \\
\sqrt{\lambda-E}\,\,{\cal J}_+(r)\, \chi_{m}
\end{pmatrix} e^{iE t}.
\end{equation}
Similarly, in the case of the minus sign the wave function reads
\begin{equation}
\Psi_-= 
\left(\frac{k}{2\lambda}\right)^{1/2}\begin{pmatrix}
-\sqrt{\lambda+E}\,{\cal J_+}(r) \,\chi_{m} \\
\sqrt{\lambda-E}\,\,{\cal J}_-(r)\, \chi_{m}
\end{pmatrix} e^{iE t}.
\end{equation}
For the
standard monopole potential (in absence of the scalar coupling), 
the lowest mode, the s-wave, has a non-vanishing probability of reaching the
core of the monopole~\cite{Kazama:1976fm}.  In our case the probability of the lowest mode reaching the core of the monopole is vanishing, since
\begin{equation}
 r^2\,\left |{\cal J}_-(r) \right|_{j=q-\frac{1}{2}}^2 \sim r^{2q}.
\end{equation}
The field itself behaves as $\Psi \sim r^{q-1}$, implying the defect operator of dimension $\Delta =q+1/2$, equal to $1$ for $q=1/2$. 

 For the construction of the propagator we will need a number 
of summation formulas for outer products between spinor spherical harmonics, i.e.\ the equivalent of eqn.~(\ref{AdditiontheoremY}) for spinors. For simplicity and for illustrational purposes
we shall restrict ourselves to the case $\bf{n}=\bf{n'}$. 
The relevant formulas for the spherical harmonics of type $\phi_{jm}^{(1)}$, $\phi_{jm}^{(2)}$ (cf.\ page~\pageref{spinorharmonics}) when $q=0$ were worked out in~\cite{Bechler:1993}.  
A numerical investigation is compatible with the following summation formulas for $q\neq 0$:
\begin{eqnarray}
\sum_m \phi^{(1)}_{jm}({\bf{n}})\phi^{(1)}_{jm}({\bf{n}})^{\dagger}&=& 
\frac{j+\frac{1}{2}}{4\pi} \, - \,\frac{q}{4 \pi}\, {\bf{n}} \cdot  \vec{{\bf {\sigma}}}, 
 \\
\sum_m \phi^{(2)}_{jm}({\bf{n}})\phi^{(2)}_{jm}({\bf{n}})^{\dagger}&=& 
\frac{j+\frac{1}{2}}{4\pi} \, + \,\frac{q}{4 \pi}\, {\bf{n}} \cdot  \vec{{\bf {\sigma}}}, \label{phi22}
 \\
\sum_m \phi^{(1)}_{jm}({\bf{n}})\phi^{(2)}_{jm}({\bf{n}})^{\dagger}&=&
-\frac{((j+\frac{1}{2})^2-q^2)^{1/2}}{4\pi} \, {\bf{n}} \cdot   \vec{{\bf {\sigma}}}, 
 \\
\sum_m \phi^{(2)}_{jm}({\bf{n}})\phi^{(1)}_{jm}({\bf{n}})^{\dagger}&=& 
-\frac{((j+\frac{1}{2})^2-q^2)^{1/2}}{4\pi} \, 
 {\bf n}  \cdot   \vec{{\bf {\sigma}}}.
\end{eqnarray}
These identities can be used to construct the Dirac propagator in the coplanar kinematics, for instance at coincident points. We will not give explicit expressions here, they have similar architecture to the derivatives of the scalar propagator, as appropriate for a fermion Green's function.

\section{One-point functions\label{one-point}}

As a first application, we study one-point functions of local operators in the presence of the 't~Hooft line:
\begin{equation}\label{1pt-gen-def}
 \left\langle \mathcal{O}(x)\right\rangle_T=
 \lim_{y\rightarrow 0}
 \frac{\left\langle T\,\mathcal{O}(x)\right\rangle}
 {\left\langle T\right\rangle|y|^{\Delta }\left\langle 
 \mathcal{O}(y)\,\mathcal{O}(0)
 \right\rangle^{\frac{1}{2}}}\,.
\end{equation}
This definition allows for an arbitrary operator normalization, and is especially convenient for non-protected operators whose normalization factor absorbs divergences and is in general scheme-dependent. It is thus important to use the same regularization scheme for the numerator and denominator.

 By scale invariance,
\begin{equation}\label{OPE}
 \left\langle \mathcal{O}(x)\right\rangle_T=\frac{\mathbbm{C}}{(2r)^\Delta }\,.
\end{equation}
The coefficient $\mathbbm{C}$ (more precisely, $2^\Delta \mathbbm{C}$) defines the weight of the unit operator in the bulk-to-defect OPE of $\mathcal{O}(r,t)$. Alternatively, applying inversion and sending the operator to infinity, we get (here assuming the unit operator normalization):
\begin{equation}
 \left\langle T(C_R)\,\mathcal{O}(x)\right\rangle\stackrel{x\rightarrow \infty }{\simeq }\frac{\mathbbm{C}R^\Delta }{|x|^{2\Delta }}\,\left\langle T(C_R)\right\rangle,
\end{equation}
where $C_R$ represents a circle of radius $R$. Thus $\mathbbm{C}R^\Delta $ is also the weight with which the operator $\mathcal{O}$ enters the OPE of a small 't~Hooft circle.

In perturbation theory, we first replace all the fields by their classical values (\ref{monopoleF}), (\ref{monopolePhi}), and then take into account quantum corrections according to (\ref{backgr-exp}), with vertices from (\ref{Ltot}) and the propagators derived in the previous section. We apply this strategy first to chiral primaries for which we were able to resum perturbative series and reproduce the exact localization results and then to non-protected operators at the one-loop level.

\subsection{Chiral primaries}

The one-point functions of chiral primaries induced by a 't~Hooft line can be obtained from the Wilson line correlators \cite{Semenoff:2001xp,Okuyama:2006jc} by combining localization with S-duality \cite{Kristjansen:2023ysz}, or directly from localization applied to the circular 't~Hooft loop
 \cite{Gomis:2011pf}. The one-point functions of chiral primaries
have been extensively studied in perturbation theory \cite{Gomis:2009ir,Gomis:2009xg,Kristjansen:2023ysz} and our
goal here is to reinterpret these results in light of the recent work on surface operators \cite{Choi:2024ktc}. One-point functions of chiral primaries for 
the surface defect of \cite{Gukov:2006jk} are polynomials in the 't~Hooft coupling  \cite{Drukker:2008wr} and are thus given by a finite number of diagrams. These diagrams were identified in  \cite{Choi:2024ktc} to be the graphs with no interaction vertices, much like for correlators of a circular Wilson loop \cite{Erickson:2000af,Drukker:2000rr,Semenoff:2001xp}. 

The one-point functions for the 't~Hooft loop, as we show here, follow a similar pattern. Perturbative series truncate at a finite order in some cases, in other cases not, but even then there is an easily identifiable polynomial part. We will show that in all cases the diagrams without internal vertices saturate the answer up to the wrapping order. This, we believe, is a universal feature of the defect one-point functions as we demonstrate by revisiting the D3-D5 system in sec.~\ref{D3D5-sec}.

\subsubsection{From Zhukovsky to Chebyshev}

The chiral primaries we consider are of the form 
\begin{equation}
 \cpo_L=\mathop{\mathrm{tr}}Z^L,\qquad Z=\Phi _1+i\Phi _2,
\end{equation}
where, just to remind, $\Phi _1$ takes on an expectation value and $\Phi _2$ does not.

The exact one-point functions of these operators can be computed with the help of  localization and are best parametrized by the Zhukovsky variables:
\begin{equation}\label{basicZhuk}
 x+\frac{1}{x}=2u,
\end{equation}
together with a quantization condition on the rapidity:
\begin{equation}\label{quantum-u}
 u=\frac{2\pi iq}{\sqrt{\lambda }}\,.
\end{equation}
The planar-exact OPE coefficients (\ref{OPE}) then have a very concise form \cite{Kristjansen:2023ysz}:
\begin{equation}\label{gen-CC}
 \CC_L=\frac{1}{i^L\sqrt{L}}\,\left[x^L\bl{+\frac{(-1)^{L+1}}{x^L}}\right].
\end{equation}
Since $x\sim 1/\sqrt{\lambda }$ at small $\lambda $, the two terms are separated by $L$ orders of perturbation theory. The second term, in blue, is thus naturally interpreted as a wrapping correction. Wrapping correction arise when all the fields in the operator start participating in quantum-mechanical interactions at the loop order equal to the length of the operator.

The Zhukovsky variables are ubiquitous in the Bethe-Ansatz equations of $\mathcal{N}=4$ SYM, from the asymptotic Bethe ansatz \cite{Beisert:2005fw} to the Quantum Spectral Curve \cite{Gromov:2013pga}. Most common normalization (the one used in \cite{Kristjansen:2023ysz}) differs from ours by an extra factor of $2\pi /\sqrt{\lambda }$ in the rapidity. The Zhukovsky cut then runs from $-\sqrt{\lambda }/2\pi $ to $\sqrt{\lambda }/2\pi $, while in our case the branch points are at $\pm 1$. The quantization condition in the conventional normalization  literally sets the rapidity to half an integer multiple of $i$. 

We should make clear that the exact one-point functions embodied in (\ref{gen-CC}) were derived for the elementary monopole \cite{Kristjansen:2023ysz}. The derivation of  (\ref{gen-CC})  is only valid  for $q=1/2$. We assume nevertheless that the large-$N$ formula continues to hold for arbitrary $q$ and in the next section provide  some circumstantial evidence for this conjecture, but this cannot replace a true derivation, and it would be very interesting to extend the localization-based analysis to an arbitrary value of magnetic charge. 

Here we list the first few OPE coefficient which hopefully illustrate the general pattern:
\begin{align}\label{CC2}
 \CC_2&=\frac{1}{\sqrt{2}}\left(\frac{4\pi q}{\sqrt{\lambda }}\right)^2
 \left(
 1+\frac{\lambda }{8\pi ^2q^2}
 \bl{
 -\frac{\lambda ^2}{128\pi ^4q^4}
 +\frac{\lambda ^3}{1024\pi ^6q^6}+\ldots 
 }
 \right),\\
 \label{CC3}
 \CC_3&=\frac{1}{\sqrt{3}}\left(\frac{4\pi q}{\sqrt{\lambda }}\right)^3
 \left(
 1+\frac{3\lambda }{16\pi ^2q^2}
 \right),\\
 \label{CC4}
 \CC_4&=\frac{1}{2}\left(\frac{4\pi q}{\sqrt{\lambda }}\right)^4
 \left(
 1+\frac{\lambda }{4\pi ^2q^2}+
 \frac{\lambda ^2}{128\pi ^4q^4}
\bl{
 -\frac{\lambda ^4}{32768\pi ^8q^8}+\ldots 
 }
 \right),\\
 \label{CC5}
  \CC_5&=\frac{1}{\sqrt{5}}\left(\frac{4\pi q}{\sqrt{\lambda }}\right)^5
 \left(
 1+\frac{5\lambda }{16\pi ^2q^2}+\frac{5\lambda ^2}{256\pi ^4q^4}
 \right),\\
 \label{CC6}
 \CC_6&=\frac{1}{\sqrt{6}}\left(\frac{4\pi q}{\sqrt{\lambda }}\right)^6
 \left(
 1+\frac{3\lambda }{8\pi ^2q^2}+
 \frac{9\lambda ^2}{256\pi ^4q^4}
 +\frac{\lambda ^3}{2048\pi ^6q^6}
\bl{
 -\frac{\lambda ^6}{8388608\pi ^{12}q^{12}}+\ldots 
 }
 \right).
\end{align}
The factor in front is just normalization from the two-point correlator in the denominator of (\ref{1pt-gen-def}). The genuine correlator with the 't~Hooft loop is the series in the brackets. Quite remarkably, for odd length the series truncate at $\lambda ^{(L-1)/2}$ signaling that only a finite number of diagrams contributes. For even length the series do not truncate but then the asymptotic part goes up to $\mathcal{O}(\lambda ^{L/2})$, then the series terminate and do not restart until the wrapping order $\mathcal{O}(\lambda ^L)$. Hence a gap between $\lambda ^{L/2}$ and $\lambda ^L$ that becomes larger and larger with growing length.

There is clearly something special about the asymptotic part, even for operators of even length. One may expect that it represents a special subset of diagrams in this case as well. We will show that the odd-length one-point functions are reproduced by the sum of diagrams without internal vertices, just like for the surface defect \cite{Choi:2024ktc}. For even length the correlator is reproduced up to the wrapping order. We leave the question of the difference between even and odd length for future investigation.

The general form of the polynomial structure follows from the relation between Zhu\-kov\-sky map and Chebyshev polynomials:
\begin{equation}\label{Zh->Ch}
 x^L+\frac{1}{x^L}=2T_L(u),\qquad 
 x^L-\frac{1}{x^L}=2\sqrt{u^2-1}\,U_{L-1}(u),
\end{equation}
for the map $x(u)$ introduced in (\ref{basicZhuk}).
Comparing to (\ref{gen-CC}) we get:
\begin{align}
 \CC_L&=\frac{2T_L(u)}{i^L\sqrt{L}},\qquad (L-{\rm odd}),
 \\
 \CC_L&=\frac{2\sqrt{u^2-1}\,U_{L-1}(u)}{i^L\sqrt{L}},\qquad (L-{\rm even}),
\end{align}
with $u$ defined by the quantization condition (\ref{quantum-u}). These are (large-$N$) exact formulas with no approximations. At odd length, $\CC_L$ is clearly polynomial and at even length it is clearly not.

In the former case, the sign alternation in (\ref{gen-CC}) leaves behind the square root. It is tempting to attribute $(-1)^{L+1}$ to a bug or a subtlety in the calculation, but there is a sanity check. Comparison to supergravity at strong coupling  \cite{Kristjansen:2023ysz} only works if the sign alternates as written. 

If we are interested in the asymptotic part, however, and are ready to ignore wrapping corrections the OPE coefficients take a simple universal form for any $L$: 
\begin{equation}\label{CPO-up-to-wrappings}
  \CC_L=\frac{2T_L(u)}{i^L\sqrt{L}}
  \bl{\,+\,{\rm wrappings}}.
\end{equation}
Indeed, the second term in (\ref{gen-CC}) is already at the wrapping order and changing the sign in front affects terms of order $\mathcal{O}(\lambda ^L)$ and beyond. We are now going to compare this expression with explicit perturbative diagrammatics.

\subsubsection{Flower diagrams}

\begin{figure}[t]
 \centerline{\includegraphics[width=10cm]{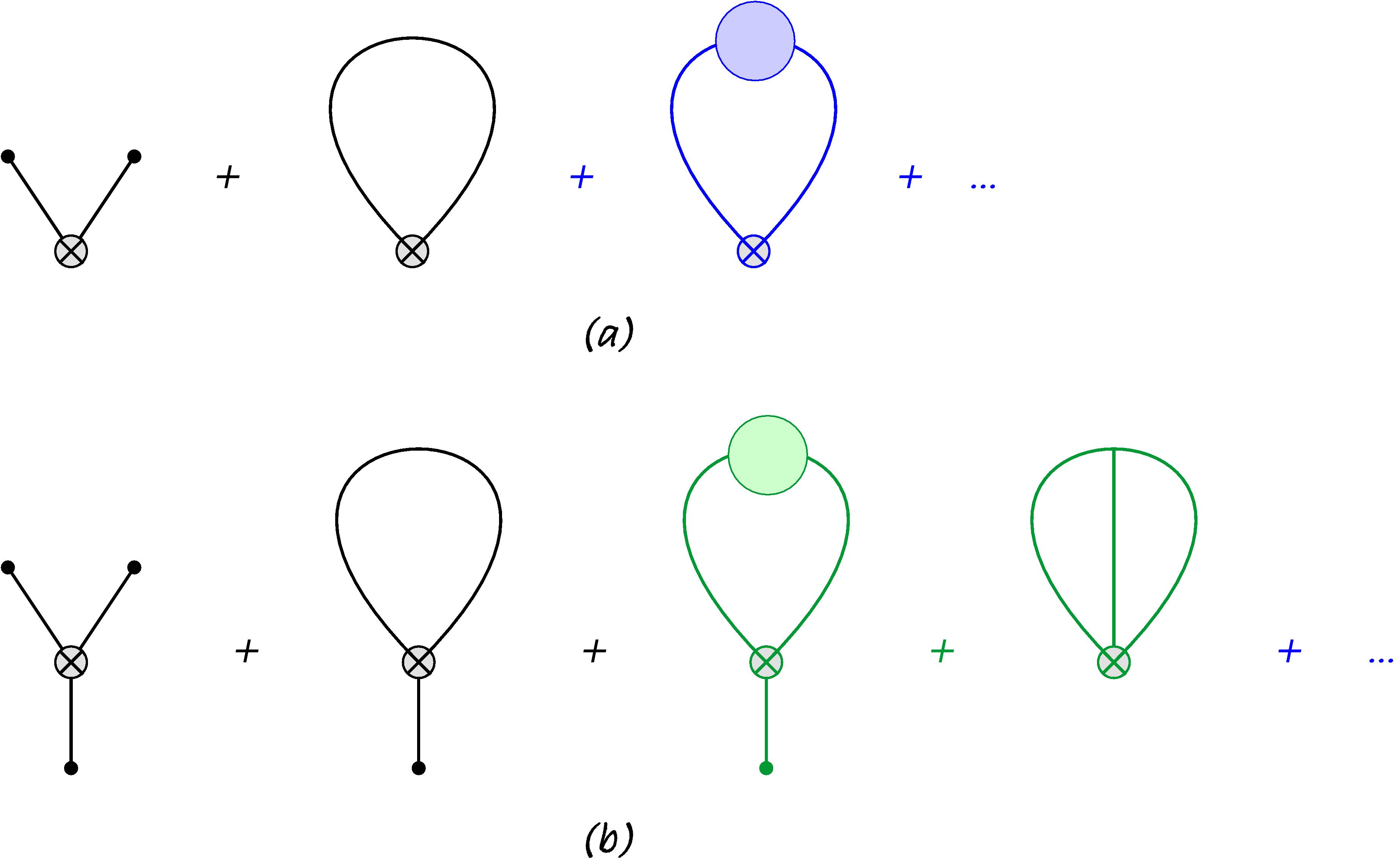}}
\caption{\label{Z2Z3}\small Perturbative expansion for (a) $\mathop{\mathrm{tr}}Z^2$, (b) $\mathop{\mathrm{tr}}Z^3$. Wrapping corrections are blue-colored. The diagrams in green mutually cancel.}
\end{figure}

An expectation value of $\mathop{\mathrm{tr}}Z^L$ can be computed order-by-order in $\lambda $ by expanding the constituent field around the classical background: $Z=q/r+z$, and computing the ensuing correlators of $z$ evaluated at coincident points.  The real and imaginary parts of $z=\phi _1+i\phi _2$  are the hard and easy fields respectively, so the propagator of $z$ is precisely the anomalous Green's function introduced in sec.~\ref{anG-sec}:
\begin{equation}
 \left\langle z^{1a}(x)z^{b1}(x')\right\rangle=\frac{\lambda \delta ^{ab}}{2N}\,G^-(x,x').
\end{equation}
Its coincidence-point limit (\ref{anom-Green}) is non-singular, so the correlation functions of $z$ are UV finite as expected. 

For example, the lowest-order diagram in fig.~\ref{Z2Z3}a gives (with the normalization restored):
\begin{equation}
 \left\langle \mathop{\mathrm{tr}}Z^2\right\rangle_T
 =\frac{4\pi ^2}{\sqrt{2}\,\lambda }\,\,\frac{q^2}{r^2}
 \left(1+\frac{\lambda }{8\pi ^2q^2}
 \bl{+\ldots }\right),
\end{equation}
which agrees with (\ref{CC2}), (\ref{OPE}).

The next order  comes from the self-energy propagator corrections (the last diagram in fig.~\ref{Z2Z3}a). This correction clearly {\it does not} vanish, because the exact asnwer  (\ref{CC2}) contains an $\mathcal{O}(\lambda ^2)$ term.
This  is  unsurprising, there is no symmetry reason for the propagator of $z$ to be protected. What is surprising is that cancellations do occur in  $\left\langle \mathop{\mathrm{tr}}Z^3\right\rangle_T$: the perturbative series in (\ref{CC3}) truncate at $\mathcal{O}(\lambda )$ and thus the complete two-loop correction must vanish. When expressed in diagrams, this means that the  triple-vertex contribution\footnote{Scalars do not have triple vertices in vacuo, but those are induced in the monopole background, see (\ref{Ltot}).} (fig.~\ref{Z2Z3}b) completely compensates the self-energy propagator diagrams.  It is easy to see that the combinatorics does not really depend on the length of the operator. Once we accept that the self-energy graphs cancel against the triple vertex for $L=3$, it should do so for any $L\geq 3$.

For $L>3$ there is also a diagram with two bubbles, at the same two-loop order. But that can be computed by easy combinatorics. The bubble must connect nearest neighbors in the trace \cite{Kristjansen:2023ysz}. At the leading order this gives a combinatorial factor of $L$. After two fields are Wick-contracted, $L-2$ vacancies remain which can be connected in $L-3$ possible ways. That gives, for arbitrary $L>2$:
\begin{equation}
 \CC_L=\frac{1}{\sqrt{L}}\left(\frac{4\pi q}{\sqrt{\lambda }}\right)^L\left[
 1+L\cdot \frac{\lambda }{16\pi ^2q^2}
 +\frac{1}{2}\,L(L-3)\cdot \left(\frac{\lambda }{16\pi ^2q^2}\right)^2
 +\ldots 
 \right].
\end{equation}
The result perfectly matches localization predictions as can be seen by inspecting  (\ref{CC3})--(\ref{CC6}).

\begin{figure}[t]
 \centerline{\includegraphics[width=3cm]{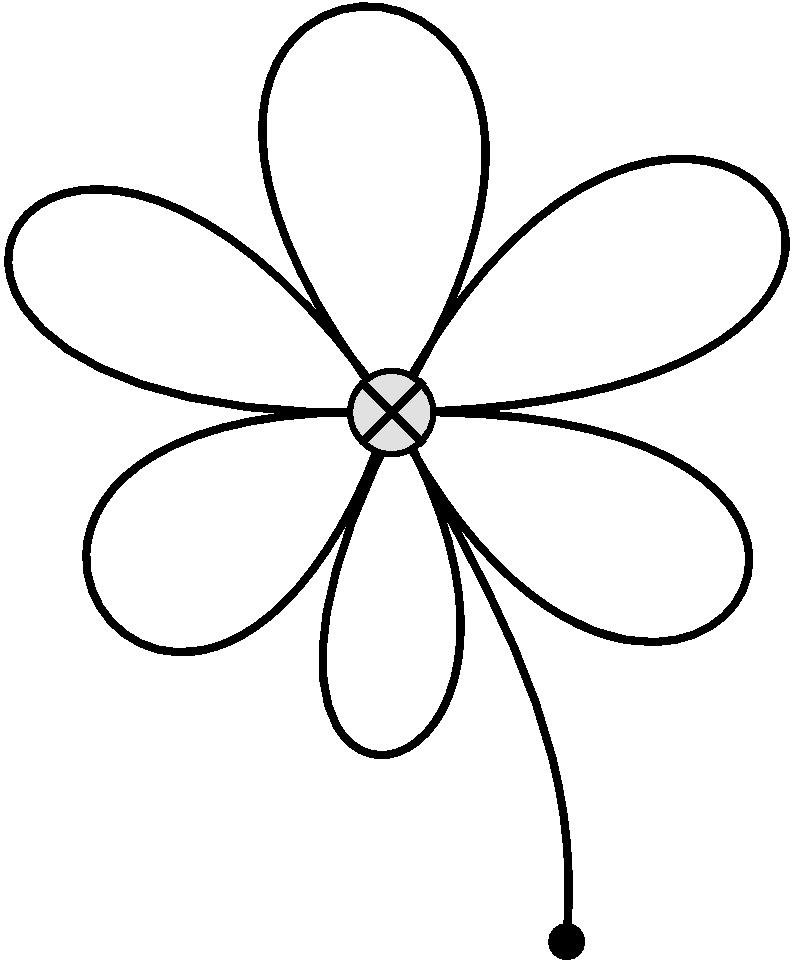}}
\caption{\label{Flower}\small A flower diagram.}
\end{figure}

We now make a crucial assumption, inspired by \cite{Choi:2024ktc}, that cancellations just observed persist to higher loops. The correlator $\left\langle \mathop{\mathrm{tr}}Z^L\right\rangle_T$ is then saturated by the diagrams with constant propagators and no interaction vertices, effectively reducing the problem to combinatorics. The diagrams at hand -- we call them flower diagrams,  fig.~\ref{Flower} --  can be resummed by reformulating the problem as a zero-dimensional field theory whose Feynman rules reproduce the requisite combinatorics.

After pulling out a factor of $(q/r)^L$, the expectation value of $Z$ is effectively set to $1$ and the propagator to $\lambda /16\pi ^2q^2$. We can thus write
\begin{equation}\label{Zfield}
 Z=\begin{bmatrix}
  1 & \varphi ^\dagger  \\ 
  \varphi  & 0 \\ 
 \end{bmatrix},
\end{equation}
with $\varphi $ a Guassian complex field with $N-1$ components and a  propagator
\begin{equation}\label{fifi}
 \left\langle \varphi ^\dagger _a\,\varphi ^b\right\rangle
 =\frac{\lambda \delta _a^b}{16\pi ^2q^2N}\,.
\end{equation}
With these definitions, the OPE coefficient of a chiral primary reduces to a correlation function in a zero-dimensional theory:
\begin{equation}\label{CPO-1d}
 \left\langle \cpo_L\right\rangle_T\equiv \frac{\CC_L}{(2r)^L}
 =\frac{1}{\sqrt{L}}\left(\frac{2\pi q}{\sqrt{\lambda }\,r}\right)^L
 \left\langle \mathop{\mathrm{tr}}Z^L\right\rangle,
\end{equation}
where the right-hand side is to be computed by the Gaussian average over $\varphi $ and $\varphi ^\dagger $.

The multiple power of the field (\ref{Zfield}) has the following generic form:
\begin{equation}\label{ZL+1}
 Z^{L+1}=\begin{bmatrix}
  P_L(\varphi ^\dagger \varphi ) & Q_L(\varphi ^\dagger \varphi ) \varphi ^\dagger  \\ 
  Q_L(\varphi ^\dagger \varphi ) \varphi  & R_L(\varphi ^\dagger \varphi ) \\ 
 \end{bmatrix},
\end{equation}
where $P_L$, $Q_L$ and $R_L$ are polynomial functions of the scalar $\varphi ^\dagger \varphi$. Multiplication by $Z$ gives rise to recursion relations:
\begin{align}
 &P_{L+1}(\xi )=P_L(\xi )+\xi Q_L(\xi ),
\nonumber \\
 &Q_{L+1}(\xi )=P_L(\xi ),
\nonumber \\
 &Q_{L+1}(\xi )=Q_L(\xi )+\xi R_L(\xi ),
\nonumber \\
 &R_{L+1}(\xi )=Q_L(\xi ).
\end{align}
Eliminating from those $P_L$ and $R_L$:
\begin{equation}
 P_L=Q_{L+1},\qquad R_L=Q_{L-1},
\end{equation}
we arrive at a recursion relation for $Q_L$ alone:
\begin{equation}
 Q_{L+1}=Q_L+\xi Q_{L-1},
\end{equation}
to be solved with the initial conditions $Q_0=1=Q_1$.

The last equation can be brought to a more familiar form by a change of variables:
\begin{equation}
 \xi =-\frac{1}{4u^2},\,\qquad Q_L=\frac{1}{(2u)^L}\,\widetilde{Q}_L.
\end{equation}
It is easy to see that
the rescaled function $\widetilde{Q}_L$ is a polynomial in $u$. The recursion relation becomes:
\begin{equation}
 \widetilde{Q}_{L+1}=2u\widetilde{Q}_L-\widetilde{Q}_{L-1},
\end{equation}
and now has the canonical form of a three-term recursion relation for orthogonal polynomials. Moreover,  the Chebyshev polynomials that we have encountered in the previous section satisfy precisely this three-term relation. For the given boundary conditions the solution is the Chebyshev polynomial of the second kind:
\begin{equation}
 \widetilde{Q}_L=U_L.
\end{equation}

From (\ref{ZL+1}) we can thus find the requisite correlator:
\begin{equation}\label{trZ-1d}
 \left\langle \mathop{\mathrm{tr}}Z^L\right\rangle=
 \left\langle 
 P_{L-1}(\varphi ^\dagger \varphi )+\varphi ^\dagger \varphi R_{L-1}(\varphi ^\dagger \varphi  )
 \right\rangle.
\end{equation}
With the explicit expressions for $P_L$, $Q_L$ and $R_L$ at hand, we can express the right-hand side through Chebyshev polynomials. Denoting, as before,
 $\varphi ^\dagger \varphi $ by $\xi$ and then replacing $\xi $ by $-1/4u^2$, we have:
\begin{equation}\label{T-from-PQ}
  P_{L-1}+\xi R_{L-1}=Q_L+\xi Q_{L-2}
  =\frac{1}{(2u)^L}\left(U_L-U_{L-2}\right)
  =\frac{2T_L}{(2u)^L}\,.
\end{equation}

The Gaussian average over $\varphi $, $\varphi ^\dagger $ follows trivially from the large-$N$ factorization:
\begin{equation}
 \left\langle \left(\varphi ^\dagger \varphi \right)^n\right\rangle\stackrel{N\rightarrow \infty }{= }
 \left\langle \varphi ^\dagger \varphi \right\rangle^n=\left(\frac{\lambda }{16\pi ^2q^2}\right)^n.
\end{equation}
The average thus amounts in replacing $\xi $ by $\lambda /16\pi ^2q^2$ or $u$ by
\begin{equation}
 u=\frac{2\pi i q}{\sqrt{\lambda }}\,,
\end{equation}
which coincides with the quantization condition (\ref{quantum-u}). It is for that reason that we used the same notation for the rapidity variable in the Zhukovsky map and the parameter in the recursion relations. 

Combining (\ref{CPO-1d}) with (\ref{trZ-1d}) and (\ref{T-from-PQ}) we finally get for the sum of  flower diagrams:
\begin{equation}
 \CC_L=\frac{2T_L(u)}{\sqrt{L}}\left(\frac{2\pi q}{\sqrt{\lambda }\,u}\right)^L
 =\frac{2T_L(u)}{i^L\sqrt{L}}\,.
\end{equation}
This coincides with the full non-perturbative answer (\ref{CPO-up-to-wrappings})  up to the wrapping order!

To conclude, combinatorics of the flower diagrams is handled by the same Chebyshev polynomials that arise from localization. The  sum of flower diagrams reproduces the full planar-exact expectation values for all chiral primaries of odd length. Operators of even length receive extra wrapping corrections, but those start at order $\mathcal{O}(\lambda ^L)$ compared to the sum of flower diagrams that truncates at $\mathcal{O}(\lambda ^{L/2})$. This explains a gap in the perturbative series observed in (\ref{CC2})--(\ref{CC6}).

\subsubsection{Remarks on D3-D5}\label{D3D5-sec}

Intriguingly, a co-dimension one defect in ${\cal N}=4$ SYM defined in terms of a Nahm pole configuration for the scalar fields show some of the same features as the 't Hooft loop studied above and the co-dimension two defect studied
in~\cite{Choi:2024ktc}. The Nahm pole configuration in question is given by the following assignment of vevs to the classical 
fields~\cite{Constable:1999ac}
\begin{equation}\label{Phiclass}
 \Phi _i^{\rm cl}=\frac{1}{x_{\perp}}\,\begin{pmatrix}
  \left(t_i\right)_{k\times k} & 0_{k\times (N-k)} \\ 
  0_{(N-k)\times k} & 0_{(N-k)\times (N-k)} \\ 
 \end{pmatrix},~i=1,2,3,\qquad \Phi ^{\rm cl}_i=0,~ i=4,5,6,
\end{equation}
where $x_{\perp}$ is the distance to the defect and where non-vanishing vevs are understood to occur only for $x_{\perp}>0$. The matrices
$t_i$, $i=1,2,3$ constitute a $k$-dimensional irreducible representation of $\mathfrak{su}(2)$. For $k=1$ the model is
to be understood as the limiting case of vanishing vevs but respectively Neumann and Dirichlet boundary conditions on the scalar fields $\Phi_i$, $i=1,2,3$ and $\Phi_i$, $i=4,5,6$~\cite{deLeeuw:2017dkd,Kristjansen:2020mhn}.
 There is no model of this type for $k=0$, see however~\cite{DeWolfe:2001pq}
for a discussion of another type of co-dimension one defect in ${\cal N}=4$ SYM.

The string theory dual of the co-dimension one system above is a D3-D5 probe brane system of Karch-Randall type~\cite{Karch:2000gx}
where a single probe D5-brane
with geometry $AdS_4\times S^2$ and $k$ units of magnetic flux on the $S^2$ shares a three-dimensional Minkowski space
with the usual stack of D3-branes~\cite{DeWolfe:2001pq,Nagasaki:2012re}. In this configuration, $k$ of the $N$ D3-branes get dissolved in the probe D5-brane.

The parameter $k$ is a tunable parameter of the defect set-up. Notice, however, that its nature is different from the
parameters of the 't Hooft line set-up. In the present paper the  we have restricted ourselves to considering a monopole
given by a single parameter $q$, the monopole charge. More general monopole charges,
parametrizing the Cartan generators of the underlying gauge group, are possible but in all cases will result in diagonal
vevs for the scalar fields~\cite{Diaconescu:1996rk}. For the D3-D5 domain wall model the vevs are generically non-Abelian.

One-point functions in the presence of the domain wall can be defined in a manner analogous to that of the 't Hooft line,
cf.\ eqn.~(\ref{1pt-gen-def}), and exhibit the same scaling behavior as in eqn.~(\ref{OPE}), just with $r$ replaced by the
transverse coordinate $x_{\perp}$. One-point functions of this set-up have been studied in depth, starting with~\cite{deLeeuw:2015hxa,Buhl-Mortensen:2015gfd}.

 Let us consider the chiral primaries 
\begin{equation}
CPO_L= \mbox{tr} Z^L,
\end{equation}
with the following field assignment
\begin{equation}
Z=\Phi_3+i\Phi_6.
\end{equation}
In this case the one-point function vanishes for odd values of $L$.
For even values of $L$ the planar  OPE coefficient $\CC_L$ can again be found by supersymmetric localization~\cite{Komatsu:2020sup} and can conveniently be expressed
using a parametrization stemming from the Zhukovsky map~\cite{Kristjansen:2020mhn}\footnote{Our normalization of the OPE coefficients in (\ref{OPE}) differs by a factor of  $2^\Delta $ from the conventions of \cite{Kristjansen:2020mhn}.}:
\begin{equation}
 \CC_L=\frac{1}{i^L\sqrt{L}}\left(
 \sum_{a=-\frac{k-1}{2}}^{\frac{k-1}{2}} x_a^L
 +k\delta _{L,2}+
\sum_{b\in \mathbbm{Z}+\frac{k-1}{2}}\frac{1}{x_b^L}
 \right),
\end{equation}
where $x_a$ is defined by
\begin{equation}
x_a+\frac{1}{x_a}=2 u_a, \hspace{0.5cm} u_a=\frac{2\pi i a}{\sqrt{\lambda}}.
\end{equation}
These are the same Zhukowski variables as in (\ref{basicZhuk}), (\ref{quantum-u}), but now the label $a$ runs over all integers or half-integers.

The OPE coefficient can be also written in a quasi-polynomial form with the help of the identities (\ref{Zh->Ch}):
 \begin{equation}
 \CC_L=\frac{2}{i^L\sqrt{L}}\left(
 \sum_{a=-\frac{k-1}{2}}^{\frac{k-1}{2}}T_L(u_a)
 +k\delta _{L,2}+
\sum_{b=\frac{k+1}{2}}^{\infty }\frac{1}{x_b^L}
 \right),
\end{equation}
clearly separating the asymptotic part (the first term) from wrapping corrections.

The sums go over half integers for even $k$ and over integers for odd $k$. It is easy to see that the wrapping corrections as expected
appear only starting from $\lambda^L$. The form of the asymptotic part  was already conjectured in~\cite{Buhl-Mortensen:2017ind}, based on 
a detailed one-loop computation involving arbitrary non-protected operators, and presented in the following equivalent form
\begin{eqnarray}\label{eq:result}
\CC_L (g)&=& \frac{2}{\sqrt{L}} \sum^{\frac{L}{2}}_{n=0} \binom{L-n}{n}\frac{L}{L-n} \frac{B_{L-2n+1}(\tfrac{1+k}{2})}{L-2n+1}
\left(\frac{\lambda}{16\pi^2}\right)^{n-L/2} \\
&&  
+\sum^{\infty}_{n=0} \frac{L[\Psi ^{(L+2 n-1)}(\frac{1+k}{2})-\Psi^{(L+2n-1)}(\frac{1-k}{2})]}{(-1)^n\, n!(L+n)!}
\left(\frac{\lambda}{16\pi^2}\right)^{n+L/2}.
\nonumber
\end{eqnarray}
Here, $B_n$ is the Bernoulli polynomial and $\Psi^{(n)}$ the $n$'th derivative of the digamma function. 
We start by noticing that the second line only starts contributing at wrapping order. Next, we will argue that the contributions in the first line come entirely from flower diagrams. 

First, the combinatorial factor $N(L,n)$ given by
\begin{equation}
N(L,n)=\binom{L-n}{n}\frac{L}{L-n},
\end{equation}
counts the number of flower diagrams with $n$ leaves which can be formed from an operator with $L$ fields. Let us place the $L$ fields inside the trace in a row. Disallowing the first and the last field to be Wick contracted with
each other one can perform $n$ nearest neighbor Wick contractions in $\binom{L-n}{n}$ ways. Furthermore, contracting
the first and the last field one can form additional $\binom{L-2-(n-1)}{n-1}=\binom{L-1-n}{n-1}$ different contractions, all
together summing up to $N(L,n)$.

The flower diagrams are generated when we compute the expectation value of $\mbox{tr} Z^L$ by expanding around the classical fields, i.e. $Z= \Phi_3^{\mbox{\footnotesize cl}}+z$ with $z=\phi_3+i\phi_6$. In the terminology used above 
the field $\phi_3$ is a hard field and $\phi_6$ is an easy one. The large-$N$ limit instructs us to Wick contract only
neighboring $z$-fields. Each such contraction generates a tadpole
(i.e.\ a leaf in the flower) and is done by
means of an anomalous propagator. Defining the anomalous propagator at coinciding points requires regularization, and
this was taken was taken care of by dimensional regularization in~\cite{Buhl-Mortensen:2016pxs,Buhl-Mortensen:2016jqo}.
The components in the \nnb-block of size $(N-k)\times (N-k)$ of the $z$-field, cf.\ eqn.~(\ref{N+k-decomp}), have  vanishing anomalous propagators. Furthermore,  the  $(N-k)\times k$ independent fields  appearing in the \knb-blocks have identical anomalous propagators. Finally, in the large-$N$ limit we can ignore the contributions to the tadpole coming from fields in the $k\times k$ block, and in this limit we find
\begin{equation} \label{z-tadpole}
\langle (z^2(x))_{a a'}\rangle= \frac{\lambda}{16\pi^2 x_{\perp}^2}\, \delta_{a,a'},  \hspace{0.5cm} a,a'=1,\ldots,k.
\end{equation}
For details we refer to~\cite{Buhl-Mortensen:2016jqo}. This constitutes another example of an anomalous 
propagator simplifying in a collinear limit, cf.\ eqn.~(\ref{tildeGeta=1}). 
Now to generate a flower diagram with $L$ fields and $n$ leaves
we should sprinkle $n$ such tadpoles among $L-2n$ classical fields each contributing a factor of $t_3/x_{\perp}$.  The delta-functions
in eqn.~(\ref{z-tadpole}) then result in a total overall factor of $\mbox{tr} \,(t_3^{L-2n})$. Picking an explicit $k$-dimensional representation of $\mathfrak{su}(2)$ one easily shows that
\begin{equation}
\mbox{tr}\, t_3^{L-2n}= \frac{2}{L-2n+1}B_{L-2n+1}\left(\frac{1+k}{2}\right).
\end{equation}
Hereby we have demonstrated that the flower diagrams account for all terms in the first line of~(\ref{eq:result}).
 
\subsection{Konishi\label{Konishi}}

Our techniques can be applied to non-protected operators as well. The tree-level one-point functions thereof, in the presence of the 't~Hooft loop, were shown to be integrable and solvable by Bethe ansatz, admitting a compact determinant representation in terms of the Bethe roots  \cite{Kristjansen:2023ysz}. The determinant formula, initially derived in the scalar sector, was extended to all possible single-trace operators of $\mathcal{N}=4$ SYM \cite{Gombor:2024api}. Moreover, a fully non-perturbative, asymptotic expression was derived in \cite{Gombor:2024api} up to an overall dressing factor. The dressing factor is the cross-channel image of the reflection phase of string excitations off the D1-brane. It can in principle be fixed by integrability bootstrap. Anticipating that the dressing phase will eventually be found we compute one-point functions of arbitrary scalar operators to the one-loop accuracy as a potential dataset to check integrability predictions. 

We start with the Konishi operator  as the simplest case :
\begin{equation}
 K=\mathop{\mathrm{tr}}\Phi _i\Phi _i.
\end{equation}
The Konishi operator is not protected and develops an anomalous dimension:
\begin{equation}\label{DeltaK}
 \Delta _K\stackrel{{\rm 1-loop}}{= }2+\frac{3\lambda }{4\pi ^2}\,.
\end{equation}
An infinite normalization factor is thus necessary to absorb the UV divergences. The normalization-independent definition (\ref{1pt-gen-def}) has a real advantage in that respect.  The UV divergences will appear in the correlator with the 't~Hooft loop, because we are dealing with the bare operator, but infinities will cancel between the numerator and denominator. It is very important to compute them in the same regularization scheme. We will use dimensional reduction where the scalar has $10-d$ components in $d$ dimensions.

The Wick contraction then gives $10-d$ terms, $9-d$  from the easy scalars and one from the hard scalar, in the terminology of sec.~\ref{BFE-sec}. Thus, to the one-loop accuracy,
\begin{equation}
 \left\langle K\,T\right\rangle\stackrel{{\rm 1-loop}}{= }\left(\frac{q}{r}\right)^2+2\,\frac{\lambda }{2}\left[
 (9-d)G(x,x)+\widetilde{G}(x,x)
 \right].
\end{equation}
The coefficient of $2$ in the second term comes from the two possible Wick contractions $\left\langle \varphi ^{1a}\varphi ^{b1}\right\rangle$ and $\left\langle \varphi ^{a1}\varphi ^{1b}\right\rangle$.

It is convenient to rewrite the hard propagator as  $\widetilde{G}=G+G^-$, then
\begin{equation}
  (9-d)G+\widetilde{G}=(10-d)G+G^-, \label{proprewrite}
\end{equation}
and using (\ref{G(x,x)}), (\ref{anom-Green})  we get:
\begin{equation}\label{<KT>}
 \left\langle K\,T\right\rangle\stackrel{{\rm 1-loop}}{= }
 \frac{q^2}{r^2}\left[
 1-\frac{\lambda }{8\pi ^2}\left(
 \frac{3}{\varepsilon }+3\ln(4\pi r^2)+4-\frac{3}{q}-\frac{1}{q^2}-6\psi (q)-3\gamma 
 \right)
 \right].
\end{equation}

The two-point function has to be computed in the same dimensional reduction scheme. The result appeared many times in the literature \cite{Beisert:2004ry}, we use the conventions of \cite{Ivanovskiy:2024vel} which consider  one-point functions on the Coulomb branch, technically a very similar problem: 
\begin{equation}\label{<KK>}
 \left\langle K(y)K(0)\right\rangle\stackrel{{\rm 1-loop}}{= }
 \frac{3\lambda ^2}{16\pi ^4y^4}\left[
 1-\frac{\lambda }{4\pi ^2}
 \left(
 \frac{3}{\varepsilon }+3\ln(\pi y^2)+4+3\gamma 
 \right)
 \right].
\end{equation}
The consistency of (\ref{1pt-gen-def}) with (\ref{DeltaK}), (\ref{<KT>}), and (\ref{<KK>}) now becomes evident: the  $1/\varepsilon $ pole in (\ref{<KT>}) and the ensuing log-correction  come out with precisely the right coefficients to match the one-loop anomalous dimension of Konishi and hence to cancel $1/\varepsilon $ from the two-point normalization. Indeed, the anomalous dimension can be read off just from the $r$-dependence of (\ref{<KT>}).  

Keeping track of the numerical factors, we find:
\begin{equation}
 \left\langle K\right\rangle_T\stackrel{{\rm 1-loop}}{= }
 \frac{4\pi ^2q^2}{\sqrt{3}\,\lambda r^{\Delta _K}}
 \left[1+\frac{\lambda }{4\pi ^2}
 \left(
 \frac{3}{2q}+\frac{1}{2q^2}+3\psi (q)-3\ln 2+3\gamma 
 \right)
 \right].
\end{equation}
Interestingly, the formal $q\rightarrow 0$ limit of this expression leaves a finite residue at one loop, that can be traced back to the contribution of the hard scalar via the anomalous Green's function. The anomalous bubble diagram  (\ref{anom-Green}) is simply independent of the magnetic charge. Of course, $q$ is quantized and varying $q$ continuously strictly speaking does not make sense.

For the minimal 't~Hooft loop with $q=1/2$ we get:
\begin{equation}
 \left\langle K\right\rangle_T\stackrel{{\rm 1-loop}}{= }
 \frac{\pi ^2}{\sqrt{3}\,\lambda r^{\Delta _K}}
 \left[1+\frac{\lambda }{4\pi ^2}
 \left(5-9\ln 2 
 \right)
 \right],\qquad \left(q={1}/{2}\right).
\end{equation}
This formula can possibly be compared with Bethe ansatz once the dressing phase in the overlap formula becomes available.

\subsection{General scalar operators}

We now turn our attention to general, non-protected, conformal operators built from scalars
\begin{equation}
\mathcal{O}=\Psi ^{I_1\ldots I_L}\mathop{\mathrm{tr}}\Phi _{I_1}\ldots 
 \Phi _{I_L}.
\end{equation}
which we refer to by means of their wavefunction $|\Psi\rangle$.
At one loop order the conformality of such an  operator is ensured by the wavefunction being an 
eigenstate of the integrable SO(6) spin chain Hamiltonian~\cite{Minahan:2002ve}.
 More precisely, at one-loop order a conformal operator and its
anomalous dimension  $\Delta^{(1)}$ obey
\begin{equation}\label{Gamma}
\Gamma |\Psi\rangle =\frac{\lambda }{16\pi ^2}\,\Delta^{(1)} |\Psi\rangle,
\end{equation}
where
\begin{equation}
 \Gamma =\frac{\lambda }{16\pi ^2}\sum_{\ell=1}^{L}\left(
 2-2P_{\ell \ell+1}+K_{\ell \ell+1}
 \right),
\end{equation}
with  $P$ and $K$  being the
 permutation and trace operators  $P^{IK}_{JL}=\delta ^I_L\delta ^K_J$, $K^{IK}_{JL}=\delta ^{IK}\delta _{JL}$
acting on neighboring sites.
To compute the one-point function of such an operator at the leading perturbative order we should simply replace the
scalar fields with their vacuum expectation values. This computation can be formulated as the computation of the
following overlap between a spin chain eigenstate and a specific boundary state~\cite{Kristjansen:2023ysz}
\begin{equation}\label{treelevel}
 \left\langle \mathcal{O}(x)\right\rangle_T=\left(\frac{2\pi ^2}{\lambda r^2}\right)^{\frac{L}{2}}
 L^{-\frac{1}{2}}\,\frac{\left\langle \mbox{Bst}\right.\!\left| \Psi \right\rangle}{\left\langle \Psi \right.\!\left|\Psi  \right\rangle^{\frac{1}{2}}}\,.
\end{equation}
where
\begin{equation}\label{Bst}
 \Bst_{I_1\ldots I_L}=n_{I_1}\ldots n_{I_L}.
\end{equation}
The tree level one-point function~(\ref{treelevel}) was found in closed form in~\cite{Kristjansen:2023ysz}. 

The strategy for computing the one-loop contribution to the general one-point functions was explained and implemented for
the D3-D5 domain wall model in~\cite{Buhl-Mortensen:2016jqo,Buhl-Mortensen:2017ind}
and for the Coulomb branch of ${\cal N}=4$ super Yang Mills in~\cite{Ivanovskiy:2024vel}. The route is
similar in the present case. There are several effects which need
to be taken into account. 
First, the operator itself gets corrected due to further mixing taking place at the two-loop
order. In particular, the scalar SO(6) sector is not closed at this loop order but involves mixing with operators containing
fermionic fields. A two-loop Feynman diagram allows three scalars to be replaced by two fermions, which leads to the curious
phenomenon of  length changing in the spin chain language. Thus, we can symbolically write our conformal eigenstate
as
\begin{equation}
|\Psi\rangle =|\Psi_B^{(0)}\rangle +\lambda\, |\Psi_B^{(1)}\rangle +\lambda\, |\Psi_F^{(0)}\rangle \equiv |\Psi_B\rangle
+\lambda \,  |\Psi_F\rangle,
\end{equation}
where the $B$ refers to a purely bosonic operator and $F$ to an operator containing also fermions (at most
two at the present order).
Secondly, the boundary state encoding $\langle T O(x)\rangle$, cf.\ eqn.~(\ref{1pt-gen-def}), gets corrected since computing
the expectation value of an operator is not any longer just a question of  inserting the classical fields but involves making Wick contractions. As explained earlier, the one-point function as defined in eqn.~(\ref{1pt-gen-def}) will be finite even for bare fields
if the numerator and denominator are calculated using the same  regularization scheme. Accordingly, we will in the following be working with the bare operators. Furthermore, for simplicity, we set $n_I=(1,\bf{0})$ as in the main text.

Making use of the concepts above we can express the numerator of 
eqn. (\ref{1pt-gen-def}) to one loop order as
\begin{eqnarray}
\lefteqn{
\langle T {\cal O}(x)\rangle/\langle T\rangle= \left(\frac{q}{r}\right)^L\left\{\langle B|\Psi_B\rangle \left(
1+\frac{\lambda}{8 \pi^2 q^2} \right)\right.} \nonumber\\
&&\left.
- \frac{\lambda}{16\pi^2}\left(\frac{1}{\epsilon} +\log (4 \pi r^2) +1 -\frac{1}{q}-2\Psi(q)-\gamma
\right) \langle B'|\Psi_B\rangle\right\},
\end{eqnarray}
where
\begin{equation}
 B'_{i_1\ldots i_L}=n_{i_1}\ldots n_{i_{l-1}} \delta_{i_l i_{l+1}} n_{i_{l+2}}\ldots n_{i_L}.
\end{equation}
Here the first line originates from a Wick contraction involving an anomalous propagator and the second line from 
Wick contractions involving propagators of easy fields, and we have made use of a rewriting similar to that of 
eqn~(\ref{proprewrite}) of section~\ref{Konishi}. The fermionic part of the wave function $|\Psi_F\rangle$ does not contribute
at this order. Now, we notice that 
\begin{equation}
\frac{\lambda}{16 \pi^2} \langle B'|=\langle B| \,\Gamma, 
\end{equation}
and thus according to eqn.~(\ref{Gamma})
\begin{eqnarray}
\lefteqn{\langle T {\cal O}(x)\rangle/ \langle T \rangle= \left(\frac{q}{r}\right)^L
\langle B|\Psi_B\rangle \left\{
1+\frac{\lambda}{8 \pi^2 q^2} \right. } \nonumber \\
&& \left.
- \left(\frac{1}{\epsilon} +\log (4 \pi r^2) +1 -\frac{1}{q}-2\Psi(q)-\gamma
\right)  \lambda \Delta^{(1)}\right\},
\end{eqnarray}
Similarly, we can express the two-point function in the denominator  of eqn. (\ref{1pt-gen-def}) as
\begin{equation}
\langle {\cal O}(y) {\cal O}(0)\rangle= L \left(\frac{\lambda}{8 \pi^2 y^2}\right)^L
\langle \Psi |\Psi\rangle \left [ 
1-2\lambda \Delta^{(1)} \left(\frac{1}{\epsilon}
+1 +\gamma + \log (\pi y^2)\right) \right],
\end{equation}
where we have exploited the fact that the wavefunction $|\Psi\rangle$ is an eigenstate of the dilatation operator $\Gamma$
with eigenvalue $\Delta^{(1)}$ and where the wave function overlap is to be understood in the following way
\begin{equation} \label{translation}
\langle \Psi |\Psi \rangle =\langle \Psi_B |\Psi_B\rangle +\frac{\lambda (L-1)}{8 \pi^2 L} \langle \Psi_F |\Psi_F\rangle,
\end{equation}
with the overlaps appearing on the right hand side being spin chain overlaps. The translation from field theory to spin
chain overlaps is slightly different for the bosonic and the fermionic part of the wave function. The fermionic operator
contains one field less than its bosonic counterpart and its two-point function involves one less propagator. The pre-factor
in~(\ref{translation}) accounts for that. We notice that when we compute the one-point function as defined by
 eqn.~(\ref{1pt-gen-def}) all divergences i.e.\ all $\epsilon$-poles cancel out. Moreover, the limiting procedure in  eqn.~(\ref{1pt-gen-def}) kills
 all $y$-dependence, and the $r$-dependence organizes itself in the way expected for an operator of one-loop conformal
 dimension $L+\lambda \Delta^{(1)}$.
 
Once the form of an operator (i.e.\ its wavefunction) is known, it is straightforward to calculate the norms and the overlaps appearing in the formula for the one-point function. This way one can easily generate data to check the predictions of integrability \cite{Gombor:2024api}. The  all-loop exact formula proposed in \cite{Gombor:2024api} contains a boundary dressing phase, currently unknown. At one loop the latter adds just one unknown coefficient and with enough data one can still do precision checks of the integrability formulas. We leave this exercise for future work.

\section{'t~Hooft line and Wilson loops\label{Wilson}}

In this section we study the correlator of the 't~Hooft line with the supersymmetric Wilson loop coupled to scalars:
\begin{equation}
 W(C)=\mathop{\mathrm{tr}}{\rm P}\exp
 \left[
 \int_{C}^{}ds\,\left(i\dot{x}^\mu A_\mu +|\dot{x}|\nu ^I\Phi _I\right)
 \right],
\end{equation}
where $\nu $ is a constant six-vector. As a contour $C$ we take a straight line a distance $r$ away from the 't~Hooft loop and parallel to it in space-time. This setup describes an interaction of a static, infinitely heavy charge with a static monopole.

By dimensional analysis, for large time extent the correlator should behave as 
\begin{equation}
 \left\langle W\,T\right\rangle\stackrel{t \rightarrow \infty }{\simeq }
 \,{\rm e}\,^{\alpha \,\frac{t }{r}}.
\end{equation}
It thus defines
the static Coulomb potential between the charge and the mo\-no\-pole:
\begin{equation}
 V(r)=-\frac{\alpha }{r}\,.
\end{equation}
The effective interaction strength $\alpha $ depends on the gauge coupling and also on the relative R-symmetry orientation of the Wilson and 't~Hooft lines:
\begin{equation}
 \nu ^In_I=\cos\varphi.
\end{equation}

The effective Coulomb charge can be expanded in a perturbative series:
\begin{equation}
 \alpha (\lambda ,\varphi ) =\sum_{n=0}^{\infty }\an(\varphi )\lambda ^n.
\end{equation}
We can use background-field perturbation theory of the 't~Hooft line to compute the coefficients order by order.
To the leading order, we just substitute the classical solution (\ref{monopolePhi}) into the  definition of the Wilson loop and find:
\begin{equation}\label{charge-tree}
 \ao=q\cos\varphi.
\end{equation}
Since the monopole field has zero electrostatic potential, the contribution comes entirely from the scalar coupling of the Wilson loop.

The tree-level interaction potential is attractive, for small angular separation, it diminishes with the growing angle and should have switched to repulsion for $\varphi >\pi /2$. This does not happen for the following reason. The monopole field resides in only one of the $N$ holonomy eigenvalues, other eigenvalues being trivial. The trace holonomy, which is the Wilson loop properly defined,
 is a sum of $N$ terms:
\begin{equation}\label{basicWT}
 \left\langle W\,T\right\rangle=\,{\rm e}\,^{t\,\frac{q }{r}\cos\varphi }
 +N-1.
\end{equation}
Because of the gauge invariance only the $t\rightarrow \infty $ limit here makes physical sense, and then only one of the two contributions survives.
The monopole contribution is dominant so long as $\cos\varphi >0$, but if $\cos\varphi <0$ the trivial eigenvalues take over and the potential identically equals zero. 

The limit $t\rightarrow \infty $, we should stress, has to be taken prior to $N\rightarrow \infty $ and has an effect of the thermodynamic limit in statistical mechanics. The competition between the two distinct,  exponentially different contributions results in a non-analytic functional form of the potential:
\begin{equation}\label{alpha0}
 \ao=\begin{cases}
 q\cos\varphi  & {\rm ~for~}\varphi <\frac{\pi }{2}
\\
 0 & {\rm ~for~}\varphi >\frac{\pi }{2}\,.
\end{cases}
\end{equation}

\subsection{One loop}

\begin{figure}[t]
 \centerline{\includegraphics[width=9cm]{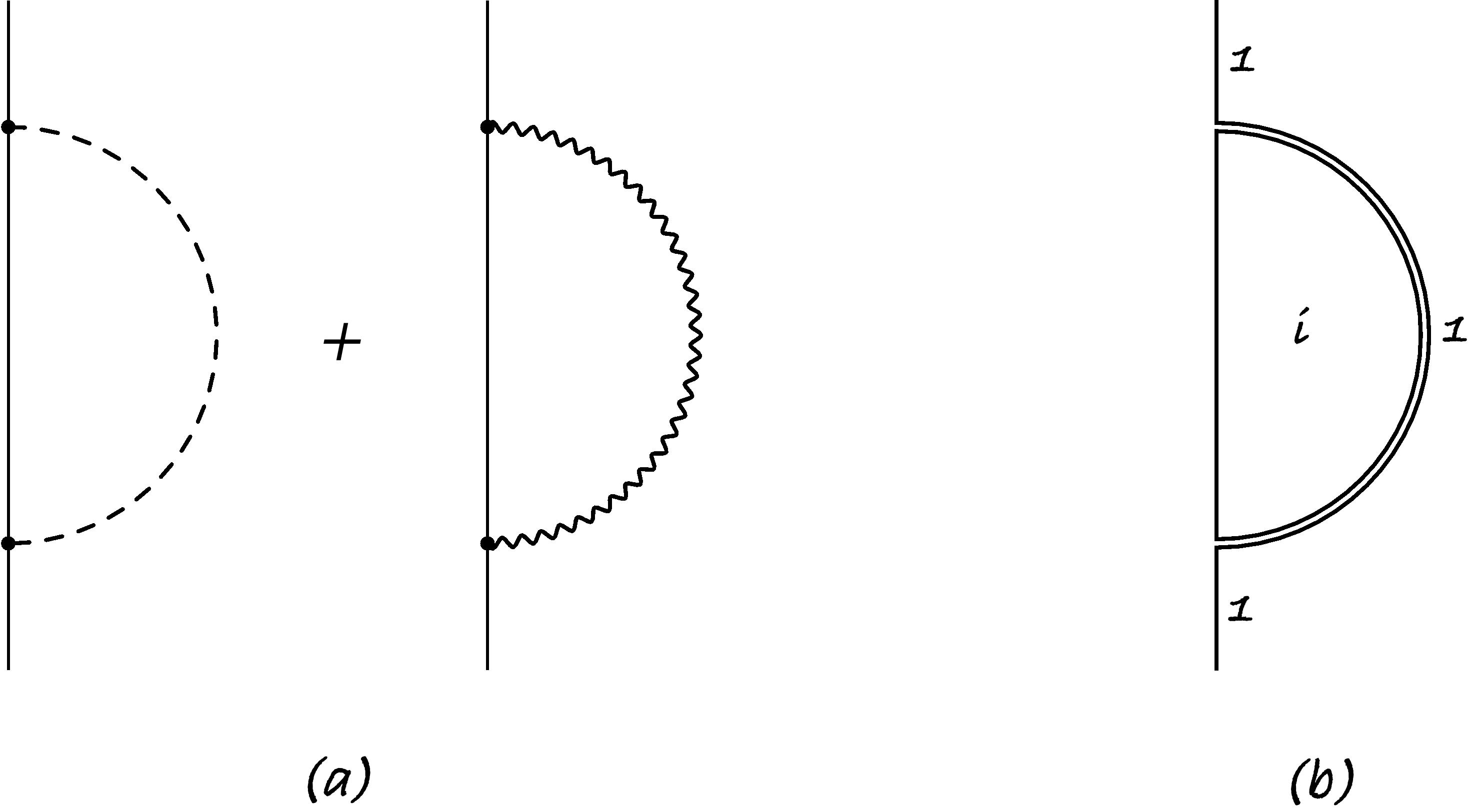}}
\caption{\label{1-loop-digrams}\small The one-loop correction to the Wilson-'t~Hooft correlator.}
\end{figure}

The next order is described by the two diagrams in fig.~\ref{1-loop-digrams}a:
\begin{eqnarray}
 \left\langle W\,T\right\rangle^{(1)}&=& \int_{0 }^{t}
 dt_2\,\int_{0}^{t_2}dt_1\,\,{\rm e}\,^{\frac{t -t_2+t_1}{r}\,q\cos\varphi }
 \left(\nu ^I\nu ^J\left\langle \Phi _I(t_1)\Phi_J (t_2)\right\rangle
 \right.
\nonumber \\
&&\vphantom{\int_{t_1}^{\tau }}
\left.
 -\left\langle A_0(t_1)A_0(t_2)\right\rangle
 \right).
\end{eqnarray}
The origin of the exponential factor lies in the color structure of the diagrams, fig.~\ref{1-loop-digrams}b.
Between the propagator endpoints the color of the Wilson line switches from $\kkb $ to $\nnb$, in the notations of (\ref{N+k-decomp}), the $\nnb$ fields do not have a classical background, and the monopole field is effectively switched off for a time interval $t_2-t_1$. As a result $t_2-t_1$ must be subtracted from $t$ in the exponent. 

The electrostatic potential is an easy field, while $\nu ^I\Phi _I$ has easy and hard components of weight $\sin\varphi $ and $\cos\varphi $, respectively. Hence,
\begin{equation}
 \left\langle \nu \Phi \,\nu \Phi \right\rangle-\left\langle A_0A_0\right\rangle
 =\frac{\lambda }{2}\,\widetilde{G}\cos^2\varphi +\frac{\lambda }{2}\,G\left(\sin^2\varphi -1\right)=\frac{\lambda }{2}\,G^-\cos^2\varphi ,
\end{equation}
where $G^-$ is the anomalous propagator (\ref{tildeGeta=1}) that contains only the lowest spherical harmonics. The Kondo effect again results in a reduction to an effective one-dimensional theory. The reason for cancellation of the infinite tower of higher partial waves is supersymmetry, in absence of the monopole the Wilson line is not renormalized at all and the one-loop correction vanishes identically.

For the effective charge we thus get:
\begin{equation}
 \al= \frac{\lambda }{2}\,r \cos^2\varphi \int_{0}^{\infty }dt\,\,{\rm e}\,^{-\alpha ^{(0)}\,\frac{\tau }{r} }\,G^-(\tau ),
\end{equation}
where the propagator is evaluated at $\eta =1$ ($\eta =\mathbf{n}\cdot \mathbf{n}'$ and the Wilson line is parallel to the 't~Hooft line so all points on it have the same $\mathbf{n}$).

The integral can be evaluated in closed form with the help of the integral representation (\ref{JJ-integral}):
\begin{equation}\label{int-rep-G-}
 G^-(\tau )=\frac{q}{8\pi r}\int_{0}^{\infty }dk\,\,{\rm e}\,^{-k\tau }\left(
  J_{q -\frac{1}{2}} (kr)^2-  J_{q +\frac{1}{2}} (kr)^2
 \right).
\end{equation}
Integrating over $\tau $ first, we get:
\begin{equation}\label{one-loop_integral}
  \alpha ^{(1)}= \frac{\lambda q\cos^2\varphi }{16\pi }
  \left(
   f_{q-\frac{1}{2}}\left(\ao\right)
   -f_{q+\frac{1}{2}}\left(\ao\right)
  \right),
\end{equation}
where
\begin{equation}\label{int-rep-f}
 f_\nu (a)=\int_{0}^{\infty }d\kappa\,\,\frac{J_\nu (\kappa )^2 }{\kappa +a}\,.
\end{equation}

The integral converges for $a>0$ and $\nu >-1/2$ (in our case $\nu \geq 0$), and evaluates to the Meijer G-function:
\begin{eqnarray}\label{functionf}
 f_\nu (a)=\frac{1}{2\pi ^{\frac{3}{2}}}\,G_{34}^{33}\left(a^2\left|
 \begin{smallmatrix} \frac{1}{2}\,\,\frac{1}{2} \\ 0\,\,\frac{1}{2}\,\,\nu \,\,-\nu  \end{smallmatrix}
 \right.\right).
\end{eqnarray}
The behavior of this function at small values of the argument is different for zero and non-zero $\nu $. When $\nu >0$,
\begin{equation}
 f_\nu (a)\stackrel{a\rightarrow 0}{\simeq }\frac{1}{2\nu }\qquad (\nu >0).
\end{equation}
and for $\nu $ exactly zero:
\begin{equation}\label{logat0}
 f_0(a)\stackrel{a\rightarrow 0}{\simeq }-\ln a.
\end{equation}
This limit  is suitable in the near-critical regime $\varphi \rightarrow \pi /2$ where the potential vanishes.

In the opposite case of large argument,
\begin{equation}\label{large-a-fnu}
 f_\nu (a)\stackrel{a\rightarrow \infty }{\simeq }\frac{1}{\pi a}\left(
 \ln(2a)-\psi \left(\nu +\frac{1}{2}\right)
 \right).
\end{equation}
We will use this result in the analysis of the fishnet limit, to be defined in the subsequent section, that makes possible an exact resummation to all orders.

As a function of the angular separation, the one-loop correction follows the same trend as the tree-level result (\ref{charge-tree}), having a maximum at $\varphi =0$, and monotonically decreasing towards $\varphi =\pi /2$ where it turns to zero. As we discussed above the potential becomes flat for $\varphi >\pi /2$ when the trivial saddle point takes over. The approach to the critical point is steeper at one loop: eq.~(\ref{one-loop_integral}) is quadratic in $\cos\varphi $ while the tree-level potential was linear. This suggests a conjecture, which we formulate for the minimal 't~Hooft loop with $q=1/2$, in which case we will be able to perform a strong-coupling check. Namely, we conjecture that the $\ell$-th loop order produces a factor of $(\cos\varphi )^\ell$ and the effective charge takes the following universal form near the critical point:
\begin{equation}\label{pi/2-scaling}
 \alpha \stackrel{\varphi \rightarrow \frac{\pi }{2}}{=}\cos\varphi \,\ha\left(\lambda \cos\varphi \right).
\end{equation}
The function $\ha$ behaves, to the one-loop accuracy, as
\begin{equation}
 \ha(x)=\frac{1}{2}-\frac{x\ln x}{32\pi }+\ldots 
\end{equation}
which follows from comparing (\ref{alpha0}) and (\ref{one-loop_integral}) with (\ref{logat0}).

It would be interesting to identify the dominant diagrams at $\varphi \rightarrow \pi /2$ and to calculate the function $\ha(x)$ exactly. We did not succeed in doing so, instead we will do an exact resummation in the opposite limit of $\cos\varphi $ becoming large. This obviously requires extending the angle $\varphi $ into the complex plane.

\subsection{Fishnet limit}

An analytic continuation in the R-symmetry angle accompanied by the limit
\begin{equation}\label{fishnet-limit}
 \varphi \rightarrow i\infty,
\end{equation}
was proposed as a means to study the non-perturbative behavior of the Wilson loop correlator  \cite{Correa:2012nk,Correa:2018lyl}. Bearing certain resemblance to the fishnet limit of the twisted $\mathcal{N}=4$ SYM \cite{Gurdogan:2015csr}, this procedure sacrifices unitarity for simplicity and leaves behind a limited set of diagrams of a certain rigid structure that can be explicitly resummed. In the Wilson loop case the survivors are the ladder diagrams \cite{Erickson:1999qv,Erickson:2000af}.
Explicit ladder resummation \cite{Erickson:1999qv,Erickson:2000af,Klebanov:2006jj,Correa:2012nk,Bykov:2012sc,Gromov:2016rrp,Beccaria:2016ejo,Beccaria:2016bfi,Cavaglia:2018lxi,Correa:2018pfn,McGovern:2019sdd} allows for comparison to string theory at strong coupling 
\cite{Correa:2012nk,Bykov:2012sc,Correa:2018lyl} and also makes direct contact with integrability of AdS/CFT via the Quantum Spectral Curve \cite{Gromov:2016rrp,Cavaglia:2018lxi,McGovern:2019sdd}. 

The ladder limit \cite{Correa:2012nk}, quite like the fishnet one  \cite{Gurdogan:2015csr}, is a double-scaling limit with the coupling going to zero simultaneously with (\ref{fishnet-limit}), such that $\lambda \,{\rm e}\,^{-i\varphi }$ remains fixed. 

Our goal is to formulate the limit (\ref{fishnet-limit}) for the correlator of the Wilson and 't~Hooft loops. To the first two orders,
\begin{equation}
 \alpha \stackrel{\varphi \rightarrow i\infty }{=}
 q\cos\varphi +\frac{\lambda \cos\varphi }{16\pi ^2q}+\ldots 
\end{equation}
where we made us of (\ref{large-a-fnu}) to simplify the one-loop contribution (\ref{one-loop_integral}). It is clear from this expression that the double-scaling limit does not make sense in our case. The two terms are of the same order if the coupling is of order one, and we will take the limit (\ref{fishnet-limit}) keeping the coupling fixed.

\begin{figure}[t]
 \centerline{\includegraphics[width=6cm]{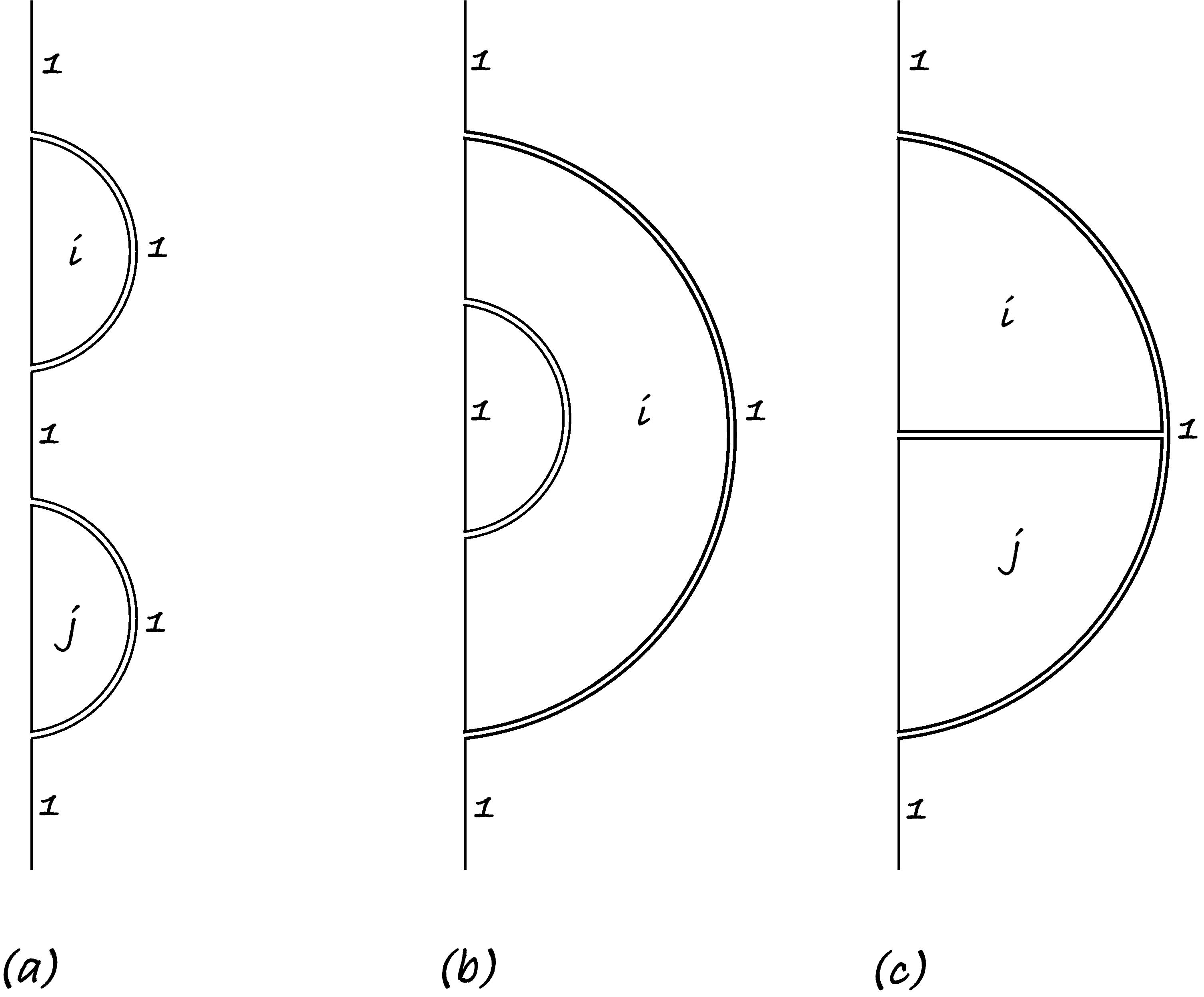}}
\caption{\label{2-l}\small Two-loop diagram topologies. The summation over the color indices  $i,j=1\ldots N-1$ is implied. The diagram (a) is an iteration of the one-loop graph. In the diagram (b) the indices have to alternate because the $\Phi _{ij}$ propagator is not affected by the external field, and with the free propagators the scalar and gluon contributions would cancel one another. The diagram (c) is subleading in the fishnet limit as explained in the text.}
\end{figure}

\subsubsection{Power counting}

To understand which diagrams are dominant at $\varphi \rightarrow i\infty $ we consider the next (two-loop) order in perturbation theory. The three topologies that may appear are shown in  fig.~\ref{2-l}. The diagram (b) is proportional to $g_{\rm YM}^4N$  and is subleading at large-$N$, the other two diagrams being of order $g_{\rm YM}^4N^2$. The dependence on the angle $\varphi $ comes from the scalar product between the R-symmetry vectors that characterize the Wilson and the 't~Hooft loop: $\nu ^In_I=\cos\varphi $. This can only arise from a propagator of the $\knb$-type field, see (\ref{N+k-decomp}), that ends on the Wilson line. The propagator has the form $D_{IJ}=n_In_J\widetilde{G}+(\delta _{IJ}-n_In_J)G$, where $G$ and $\widetilde{G}$ are easy and hard Green's functions. The one-loop correction in fig.~\ref{1-loop-digrams}
is proportional to $\nu _ID_{IJ}\nu _J\sim G^-\cos^2\varphi $, in accord with the explicit calculation. The diagram (a) in fig.~\ref{2-l} is thus proportional to $\cos^4\varphi $, while for  (c) the algebra gives $\nu _ID_{IJ}D_{JK}\nu _K\sim G^-G^+\cos^2\varphi $, two powers of $\cos\varphi $ short of the diagram (a). 

\subsubsection{Summing up ladders}

\begin{figure}[t]
 \centerline{\includegraphics[width=12cm]{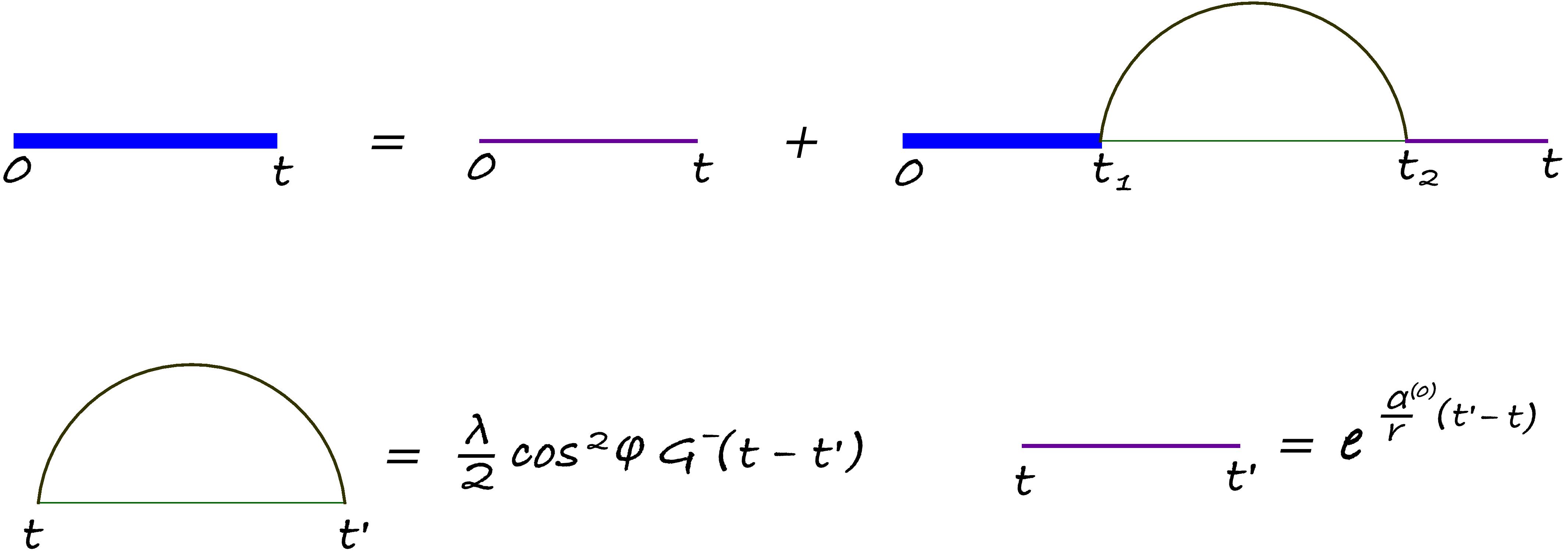}}
\caption{\label{dys}\small Pictorial form of the Dyson equation (\ref{Dyson-eq}). Its iterative solution generates diagrams carrying the largest power of $\cos \varphi $ at each order in perturbation theory.}
\end{figure}

It should be clear from the above that the leading contribution comes from the iteration of the one-loop diagram in fig.~\ref{1-loop-digrams}. The Dyson equation that resums these diagrams is shown graphically in fig.~\ref{dys}. In the analytic form,
\begin{equation}
 \label{Dyson-eq}
 W(t)=\,{\rm e}\,^{\frac{\ao}{r}\,t}+\frac{\lambda }{2}\,\cos^2\varphi \int_{0}^{t}dt_2\,
 \int_{0}^{t_2}dt_1\,
 W(t_1)G^-(t_2-t_1)\,{\rm e}\,^{\frac{\ao}{r}\,(t-t_2)},
\end{equation}
where the zeroth-order Coulomb charge is given by (\ref{charge-tree}) and $G^-(t)$ is the anomalous propagator (\ref{int-rep-G-}).

This equation embodies the Kondo effect we discussed before: computation of the correlator reduces to a one-dimensional effective theory with the propagator containing only a single partial wave of the original KK reduction on $S^2$.
The Dyson equation still looks quite complicated but in fact can be explicitly solved. It carries certain similarity to the loop equation for the Hermitian one-matrix model \cite{Wadia:1980rb,Migdal:1983qrz}. The diagrams it resums structurally resemble those that describe an RG flow on a line defect generated by generalized free fields \cite{Nagar:2024mjz}. As in these two cases, the equation can be solved by the Laplace transform \cite{Migdal:1983qrz}.

First, we act on both sides with a differential operator
$$
 r\,{\rm e}\,^{\frac{\ao}{r}\,t}\,\frac{d}{dt}\,\,{\rm e}\,^{-\frac{\ao}{r}\,t}
 =r\frac{d}{dt}-\ao.
$$
That gives an integro-differential equation
\begin{equation}
 r\,\frac{dW(t)}{dt}=\ao W(t)+\frac{\lambda r}{2}\,\cos^2\varphi \int_{0}^{t}dt_1\,W(t_1)G^-(t-t_1),
\end{equation}
to be solved with the  boundary condition
\begin{equation}
 W(0)=1.
\end{equation}

The non-locality is now contained in a convolution integral, to solve it we apply the Laplace transform:
\begin{equation}
 \widehat{W}(s)=\int_{0}^{\infty }dt\,\,{\rm e}\,^{-\frac{s}{r}\,t}\,W(t),
 \qquad 
 \widehat{D}(s)=\frac{\lambda r}{2}\,\cos^2\varphi \int_{0}^{\infty }dt\,\,{\rm e}\,^{-\frac{s}{r}\,t}\,G^-(t).
\end{equation}
The convolution then becomes a product, and the Dyson equation takes on a simple algebraic form:
\begin{equation}
 s\widehat{W}(s)-r=\ao \widehat{W}(s)+\widehat{D}(s)\widehat{W}(s),
\end{equation}
with the solution:
\begin{equation}\label{hatW}
 \widehat{W}(s)=\frac{r}{s-\ao-\widehat{D}(s)}\,.
\end{equation}

The Laplace image of (\ref{int-rep-G-}) is expressed through the function $f_\nu (a)$ defined in (\ref{int-rep-f}), (\ref{functionf}):
\begin{equation}
 \widehat{D}(s)=\frac{\lambda q\cos^2\varphi }{16\pi }\left(
 f_{q-\frac{1}{2}}(s)-f_{q+\frac{1}{2}}(s)
 \right).
\end{equation}
Since $f_\nu (s)$ is a fairly complicated function the inverse Laplace transform of $\widehat{W}(s)$ cannot be evaluated in the closed form. But the function $W(t)$ itself is an auxiliary object, not even gauge invariant at finite $t$. What we are really interested in is its large-time asymptotics which defines the potential:
\begin{equation}
 W(t)\stackrel{t\rightarrow \infty }{\simeq }\,{\rm e}\,^{\frac{\alpha }{r}\,t}.
\end{equation}

An exponential growth of a function is governed by the rightmost singularity of its Laplace image in the complex plane.  It is quite obvious from the exact solution (\ref{hatW}) that this singularity is a pole determined as a solution of the transcendental equation 
\begin{equation}
 \alpha =\ao+\widehat{D}(\alpha ).
\end{equation}
This gives:
\begin{equation}\label{main-alpha}
 \alpha =q\cos\varphi +\frac{\lambda q\cos^2\varphi }{16\pi }\left(
 f_{q-\frac{1}{2}}(\alpha )-f_{q+\frac{1}{2}}(\alpha )
 \right).
\end{equation}
where we used  (\ref{charge-tree}) for  $\ao$. 

This is the equation that resums ladders for the Wilson-'t~Hooft correlator. At finite $\lambda $ and $\cos\varphi $ it cannot be further simplified.
An iterative solution generates ladder diagrams order by order, and indeed the one-loop correction (\ref{one-loop_integral}) is reproduced at the first iteration.
At this order the result is actually exact, but starting with two loops ladders are not the only diagrams that contribute. 

For consistency, to emphasize ladders, we should take the limit $\cos\varphi \rightarrow \infty $. The Coulomb charge  is then large as evident from the tree-level approximation. Consequently we can use the asymptotic large-argument approximation (\ref{large-a-fnu}) for $f_\nu (\alpha )$. Then,\begin{equation}
  \alpha =q\cos\varphi +\frac{\lambda \cos^2\varphi }{16\pi ^2\alpha }\,.
\end{equation}

Solving for $\alpha  $ we get:
\begin{equation}\label{alpha-ladders}
 \alpha =\frac{q\cos\varphi }{2}\left(1+\sqrt{1+\frac{\lambda }{4\pi ^2q^2}}\right).
\end{equation}
This remarkably simple expression is our final result. It describes the fully non-perturbative correlator of the Wilson loop with the 't~Hooft line in the fishnet limit (\ref{fishnet-limit}).  Let us note that there exists another configuration with the Wilson loop linking the 't Hooft line where the correlator can be
computed exactly using localization~\cite{Giombi:2009ek}.

For the minimal monopole charge $q=1/2$ the effective Coulomb coupling (\ref{alpha-ladders}) coincides with the energy of the giant magnon  at the edge of the Brillouin zone  \cite{Beisert:2004hm,Beisert:2005tm}. We do not known if a deeper connection to the AdS/CFT integrability can be established, but a simple consequence of this observation is that the minimal Wilson-'t~Hooft correlator shares the same radius of convergence $\lambda _{\rm c}=\pi ^2$ with  many other correlation functions calculable by integrability or localization (see \cite{Russo:2013kea}, for an example).

At strong coupling the Coulomb charge becomes independent of $q$:
\begin{equation}\label{strong-ladder}
 \alpha \stackrel{\lambda \rightarrow \infty }{\simeq }\frac{\sqrt{\lambda }\cos\varphi }{4\pi }\,,
\end{equation}
a very suggestive result. Indeed,  in holography the correlator is represented by a string stretched between the Wilson loop and a stack of D1-branes. The area of the string world-sheet (and with it the leading semi-classical approximation to the correlator) should not depend on the number of D-branes in the stack. Somewhat oversimplifying, multiple D-branes produce an overall combinatorial prefactor but cannot  affect the classical string solution itself. 

The square-root scaling with $\lambda $ is  also consistent with holography. It is natural to expect as well that the area of the world-sheet  grows with $\cos\varphi $ when it is taken  to be large. We will construct the string solution explicitly in the next section and compare the string predictions with (\ref{strong-ladder}) including numerical prefactors.

\subsection{Strong coupling}

The holographic Wilson-'t~Hooft correlator was studied in  \cite{Minahan:1998xb,Gorsky:2009pc} in the regime of comparable 
D1-brane and string tensions. In our case the D1-brane can be regarded as infinitely heavy, its tension scaling linearly with $N$. But we need to switch on a different parameter to model a relative R-symmetry orientation of the Wilson and 't~Hooft loops.  

The 't~Hooft line in the absence of backreaction is described by a planar D1-brane embedded vertically in $AdS_5$. The string is attached by one end to the brane while the other end approaches the AdS boundary along the Wilson line. The standard Dirichlet-Neumann boundary conditions require the string worldsheet to meet the D1-brane at the right angle. 

This setup corresponds to the minimal magnetic charge $q=1/2$, otherwise one should consider a stack of $2q$ D1-branes. But at the classical level the collective behavior of the D1-branes should  not matter and the leading strong-coupling result should be independent of the magnetic charge. 

\begin{figure}[t]
 \centerline{\includegraphics[width=10cm]{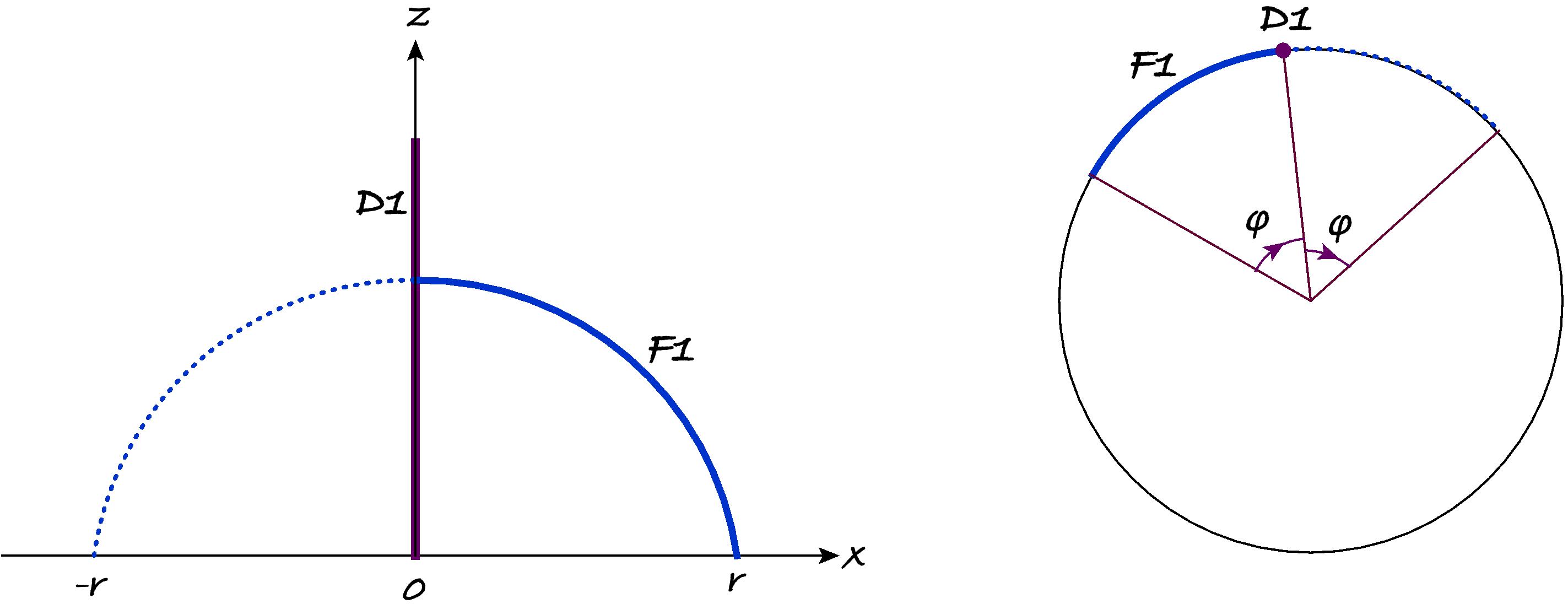}}
\caption{\label{strong-figure}\small The minimal surface describing the correlator of Wilson and 't~Hooft lines at strong coupling. On the left: the time slice of the string worldsheet; on the right: the string trajectory along the big circle of $S^5$. The solution is obtained by placing a mirror charge behind the monopole and cutting the minimal surface by half.}
\end{figure}

The requisite string solution can be constructed by the method of images.
Once a virtual charge is placed behind the monopole, as shown in fig.~\ref{strong-figure}, the minimal surface connecting the two Wilson lines meets the D1-brane at the right angle as prescribed by the boundary conditions, and thus half of the worldsheet describes the Wilson-'t~Hooft correlator. The string solution for the Wilson lines is known \cite{Maldacena:1998im}, its area can be expressed as an indirect function of the angle in terms of the elliptic integrals $\EE(m)$ and $\KK(m)$:
\begin{eqnarray}
\label{eff-charge-strong}
 \alpha &=&\frac{\sqrt{\lambda }}{2\pi }\,\,\frac{\left[\EE-(1-m)\KK\right]^2}{\sqrt{m(1-m)}},
\\
\varphi &=&\sqrt{1-2m}\,\KK.
\end{eqnarray}
Here $m$ is the elliptic modulus and we have taken into account the various factors of two arising upon cutting the minimal surface in two halves.

The angle is zero when $m=1/2$, which gives:
\begin{equation}
 \alpha (0)=\frac{\sqrt{\lambda }\,\Gamma \left(\frac{3}{4}\right)^4}{4\pi ^2}\,.
\end{equation}
This agrees with the calculation of~\cite{Minahan:1998xb} once the D1-brane tension there is taken much larger than that of the string. 

\begin{figure}[t]
 \centerline{\includegraphics[width=6cm]{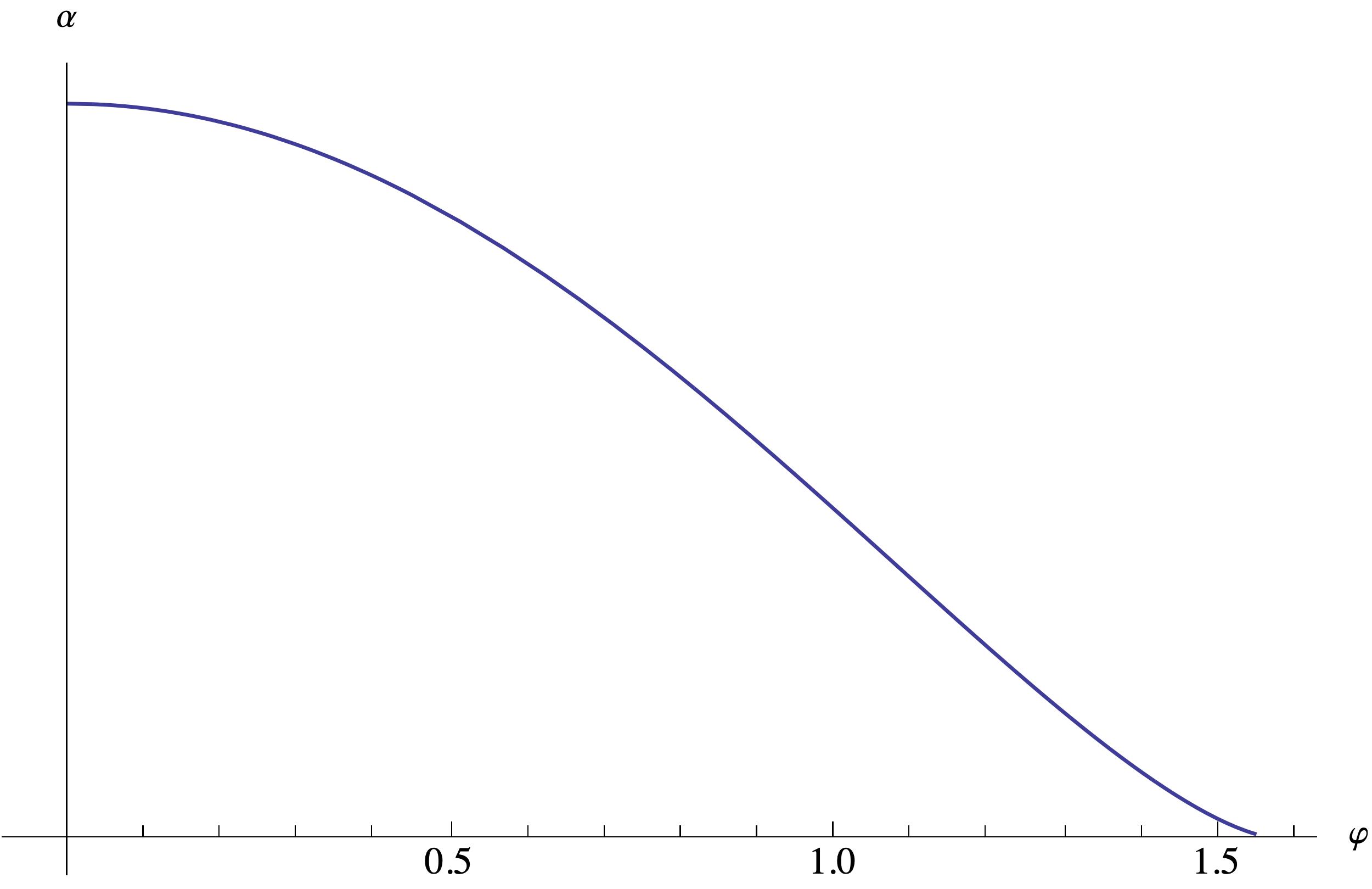}}
\caption{\label{alpha-plot}\small The Coulomb charge at strong coupling, as a function of the R-symmetry angle.}
\end{figure}

When $m$ varies between $1/2$ and $0$ the angle grows from zero to $\pi /2$  as shown in fig.~\ref{alpha-plot}. At $\varphi =\pi /2$ the Coulomb charge turns to zero\footnote{For two Wilson loops the area vanishes at $\varphi =\pi $ \cite{Maldacena:1998im}, here the critical angle is twice as small because the worldsheet is divided by half.}, just like at weak coupling. Close to the critical point:
\begin{equation}
 \alpha \stackrel{\varphi \rightarrow \frac{\pi }{2}}{\simeq }
 \sqrt{\frac{\lambda \cos^3\varphi }{54\pi }}\,.
\end{equation}
This is consistent with the conjectured scaling (\ref{pi/2-scaling}), and defines the strong-coupling asymptotics of the scaling function introduced there:
\begin{equation}
 \ha (x)\stackrel{x\rightarrow \infty }{\simeq }\sqrt{\frac{x}{54\pi }}\,.
\end{equation}
It would be very interesting to calculate the whole function non-per\-tur\-ba\-tive\-ly.

For $\varphi >\pi /2$ the connected minimal surfaces ceases to exist. The dominant saddle-point is then the trivial $AdS_2$ solution stretching to the horizon instead of ending on the D1-brane. The competition between connected and disconnected solutions is the strong-coupling counterpart of the interplay between the monopole and the trivial background in (\ref{basicWT}). The disconnected solution has a disc topology while the connected minimal surface is a cylinder (this is perhaps easier to visualize for a closed Wilson loop, e.g. a circle). The disconnected contribution hence is $\mathcal{O}(1/g_{\rm s})=\mathcal{O}(N)$ compared to $\mathcal{O}(1)$ for the connected surface, again in accord with the weak-coupling intuition.

Analytic continuation to pure imaginary angles corresponds to $m$ varying between $1/2$ and $1$.  The extreme case of $\varphi \rightarrow i\infty $ is attained when $m\rightarrow 1^-$. The angle then logarithmically diverges:
\begin{equation}
 \varphi \stackrel{m\rightarrow 1}{\simeq }\frac{i}{2}\,\ln\frac{16}{1-m}\,,
\end{equation}
 while the area grows as a power:
\begin{equation}
 \alpha \stackrel{m\rightarrow 1}{\simeq }\frac{\sqrt{\lambda }}{8\pi }\,\,\frac{4}{\sqrt{1-m}}\simeq \frac{\sqrt{\lambda }}{8\pi }\,\,{\rm e}\,^{-i\varphi }\simeq \frac{\sqrt{\lambda }\,\cos\varphi }{4\pi }\,,
\end{equation}
and this agrees exactly with the result of ladder resummation, eq.~(\ref{strong-ladder}).

\section{Conclusions \label{conclusion}}

We developed a perturbative framework for the supersymmetric 't~Hooft loop in  $\mathcal{N}=4$ super-Yang-Mills theory. Albeit we mainly focussed on the one-loop calculations (and simple iterations thereof), our quantization of the monopole enables higher loop perturbative computations such as those in~\cite{Pufu:2013vpa}. 

Supersymmetry tangibly simplifies the problem, for instance, all modes come out integer or half-integer-valued. This is not case already for the vector modes of the conventional, non-supersymmetric monopole \cite{Olsen:1990jm}. We expect similar simplifications to occur for supersymmetric monopoles in ABJM theory. In three dimensions a monopole represents a local operator \cite{Borokhov:2002ib,Borokhov:2002cg,Grignani:2007xz}, so many of the two-point functions that we studied here in ABJM will be forbidden by conformal symmetry, but some may be non-trivial, for instance correlators with the Wilson loops.

Another consequence of supersymmetry is absence of the Callan-Rubakov effect. As discussed in sec.~\ref{Fermions}, the supersymmetric 't~Hooft loop does not carry fermion zero modes, and fermion fields  satisfy conventional Dirichlet boundary conditions on the defect.

The 't Hooft line is amenable to practically all the exact methods available, integrability and localization being prime examples. One could imagine the boundary conformal
bootstrap also having a role to play. We concentrated on the bulk correlation functions in this paper, it should be possible to study the operator spectrum and correlation functions on the defect itself using the same perturbative techniques, and in perspective also by integrability and bootstrap. 

\subsection*{Acknowledgements}
We are grateful to J.\ Ambj\o rn, J.~Gomis, J.~Minahan and A.~Sever for interesting discussions. 
We gratefully acknowledge support from the Simons Center for Geometry and Physics, Stony Brook University during the workshop ``Exact approaches to low-supersymmetry AdS/CFT."
 K.Z.\ would like to thank Tel Aviv University and IIP, Natal for hospitality during the course of this work, and the Simons Foundation for the support during the visit to IIP via award 1023171-RC.
This work was supported by  DFF-FNU through grant number 1026-00103B (C.K.) and by VR grant 2021-04578 (K.Z.). 
Nordita was partially supported by Nordforsk. 

\appendix

\section{Partial-wave expansion}
\label{part-wave-app}

For $q=0$ (no monopole), the partial-wave representation (\ref{part-wave}) should be equivalent to the standard coordinate-space propagator. Here we show that this is indeed the case. Setting $q=0$ in  (\ref{part-wave})  we have:
\begin{equation}
 G(x,x')=\frac{1}{8\pi^2 rr'}
 \sum_{j}^{}(2j+1)P_{j}^{}(\eta )Q_{j}(\xi ),
\end{equation}
where now $P_j$ are the standard Legendre polynomials and the $AdS_2$ propagator is expressed through the Legendre function of the second kind $Q_j$, in virtue of (\ref{AdS2}).

Using a summation formula for the Legendre functions \cite{gradshteyn2014table}
\begin{equation}
 \sum_{j=0}^{\infty }(2j+1)P_{j}^{}(\eta )Q_{j}(\xi )=\frac{1}{\eta -\xi }\,,
\end{equation}
and re-expressing the cross ratios (\ref{crossAdS2}), (\ref{crossS2}) through coordinates, we get
\begin{equation}
 G(x,x')=\frac{1}{8\pi^2 rr'}\,\,
 \frac{1}{\frac{\left(t-t'\right)^2+r^2+r'^2}{2rr'}-\mathbf{n}\cdot \mathbf{n}'}=
 \frac{1}{4\pi ^2\left(x-x'\right)^2}\,,
\end{equation}
as advertised.

%\bibliographystyle{nb}
%\bibliography{refs-TL}

\begin{thebibliography}{10}
\ifx\href\asklfhas\newcommand{\href}[2]{#2}\fi
\raggedright
\small
\parskip 0pt

\bibitem{Pestun:2016zxk}
V.~Pestun et~al.,
\textit{``{Localization techniques in quantum field theories}''},
\textsf{J.~Phys.~A~50,~440301~(2017)},
\href{http://arXiv.org/abs/1608.02952}{\texttt{1608.02952}}.
%
\bibitem{Liendo:2012hy}
P.~Liendo, L.~Rastelli and B.~C.~van~Rees,
\textit{``{The Bootstrap Program for Boundary CFT\_d}''},
\textsf{JHEP~1307,~113~(2013)},
\href{http://arXiv.org/abs/1210.4258}{\texttt{1210.4258}}.
%
\bibitem{Kristjansen:2024dnm}
C.~Kristjansen and K.~Zarembo,
\textit{``{Integrable Holographic Defect CFTs}''},
\href{http://arXiv.org/abs/2401.17144}{\texttt{2401.17144}}.
%
%%CITATION = HEP-TH/0003055;%%
\bibitem{Erickson:2000af}
J.~K.~Erickson, G.~W.~Semenoff and K.~Zarembo,
\textit{``{Wilson loops in N = 4 supersymmetric Yang-Mills theory}''},
\textsf{Nucl.~Phys.~B582,~155~(2000)},
\href{http://arXiv.org/abs/hep-th/0003055}{\texttt{hep-th/0003055}}.
%
\bibitem{Buhl-Mortensen:2016pxs}
I.~Buhl-Mortensen, M.~de~Leeuw, A.~C.~Ipsen, C.~Kristjansen and M.~Wilhelm,
\textit{``{One-loop one-point functions in gauge-gravity dualities with
  defects}''},
\textsf{Phys.~Rev.~Lett.~117,~231603~(2016)},
\href{http://arXiv.org/abs/1606.01886}{\texttt{1606.01886}}.
%
%%CITATION = ARXIV:1611.04603;%%
\bibitem{Buhl-Mortensen:2016jqo}
I.~Buhl-Mortensen, M.~de~Leeuw, A.~C.~Ipsen, C.~Kristjansen and M.~Wilhelm,
\textit{``{A Quantum Check of AdS/dCFT}''},
\textsf{JHEP~1701,~098~(2017)},
\href{http://arXiv.org/abs/1611.04603}{\texttt{1611.04603}}.
%
\bibitem{GimenezGrau:2018jyp}
A.~Gimenez~Grau, C.~Kristjansen, M.~Volk and M.~Wilhelm,
\textit{``{A Quantum Check of Non-Supersymmetric AdS/dCFT}''},
\textsf{JHEP~1901,~007~(2019)},
\href{http://arXiv.org/abs/1810.11463}{\texttt{1810.11463}}.
%
\bibitem{Gimenez-Grau:2019fld}
A.~Gimenez-Grau, C.~Kristjansen, M.~Volk and M.~Wilhelm,
\textit{``{A Quantum Framework for AdS/dCFT through Fuzzy Spherical Harmonics
  on $S^4$}''},
\textsf{JHEP~2004,~132~(2020)},
\href{http://arXiv.org/abs/1912.02468}{\texttt{1912.02468}}.
%
\bibitem{Choi:2024ktc}
C.~Choi, J.~Gomis and R.~I.~Garc\'\i{}a,
\textit{``{Surface operators and exact holography}''},
\textsf{JHEP~2412,~195~(2024)},
\href{http://arXiv.org/abs/2406.08541}{\texttt{2406.08541}}.
%
\bibitem{deLeeuw:2024qki}
M.~de~Leeuw and A.~Holguin,
\textit{``{Integrable Conformal Defects in N=4 SYM}''},
\href{http://arXiv.org/abs/2406.13741}{\texttt{2406.13741}}.
%
\bibitem{Kristjansen:2023ysz}
C.~Kristjansen and K.~Zarembo,
\textit{``{\textquoteright{}t Hooft loops and integrability}''},
\textsf{JHEP~2308,~184~(2023)},
\href{http://arXiv.org/abs/2305.03649}{\texttt{2305.03649}}.
%
\bibitem{Gombor:2024api}
T.~Gombor and Z.~Bajnok,
\textit{``{Dual overlaps and finite coupling \textquoteright{}t Hooft
  loops}''},
\textsf{JHEP~2412,~034~(2024)},
\href{http://arXiv.org/abs/2408.14901}{\texttt{2408.14901}}.
%
\bibitem{Brennan:2023tae}
T.~D.~Brennan,
\textit{``{A new solution to the Callan Rubakov effect}''},
\textsf{JHEP~2411,~170~(2024)},
\href{http://arXiv.org/abs/2309.00680}{\texttt{2309.00680}}.
%
\bibitem{vanBeest:2023dbu}
M.~van~Beest, P.~Boyle~Smith, D.~Delmastro, Z.~Komargodski and D.~Tong,
\textit{``{Monopoles, Scattering, and Generalized Symmetries}''},
\href{http://arXiv.org/abs/2306.07318}{\texttt{2306.07318}}.
%
\bibitem{vanBeest:2023mbs}
M.~van~Beest, P.~Boyle~Smith, D.~Delmastro, R.~Mouland and D.~Tong,
\textit{``{Fermion-monopole scattering in the Standard Model}''},
\textsf{JHEP~2408,~004~(2024)},
\href{http://arXiv.org/abs/2312.17746}{\texttt{2312.17746}}.
%
\bibitem{Khoze:2023kiu}
V.~V.~Khoze,
\textit{``{Scattering amplitudes of fermions on monopoles}''},
\textsf{JHEP~2311,~214~(2023)},
\href{http://arXiv.org/abs/2308.09401}{\texttt{2308.09401}}.
%
\bibitem{Buhl-Mortensen:2017ind}
I.~Buhl-Mortensen, M.~de~Leeuw, A.~C.~Ipsen, C.~Kristjansen and M.~Wilhelm,
\textit{``{Asymptotic One-Point Functions in Gauge-String Duality with
  Defects}''},
\textsf{Phys.~Rev.~Lett.~119,~261604~(2017)},
\href{http://arXiv.org/abs/1704.07386}{\texttt{1704.07386}}.
%
%%CITATION = ARXIV:1512.06704;%%
\bibitem{Gurdogan:2015csr}
{\"O}.~G{\"u}rdo{\u g}an and V.~Kazakov,
\textit{``{New Integrable 4D Quantum Field Theories from Strongly Deformed
  Planar $\mathcal N = $ 4 Supersymmetric Yang-Mills Theory}''},
\textsf{Phys.~Rev.~Lett.~117,~201602~(2016)},
\href{http://arXiv.org/abs/1512.06704}{\texttt{1512.06704}}.
%
\bibitem{Minahan:1998xb}
J.~A.~Minahan,
\textit{``{Quark - monopole potentials in large N superYang-Mills}''},
\textsf{Adv.~Theor.~Math.~Phys.~2,~559~(1998)},
\href{http://arXiv.org/abs/hep-th/9803111}{\texttt{hep-th/9803111}}.
%
\bibitem{Gorsky:2009pc}
A.~Gorsky, A.~Monin and A.~V.~Zayakin,
\textit{``{Correlator of Wilson and t'Hooft Loops at Strong Coupling in N=4 SYM
  Theory}''},
\textsf{Phys.~Lett.~B~679,~529~(2009)},
\href{http://arXiv.org/abs/0904.3665}{\texttt{0904.3665}}.
%
\bibitem{tHooft:1977nqb}
G.~'t~Hooft,
\textit{``{On the Phase Transition Towards Permanent Quark Confinement}''},
\textsf{Nucl.~Phys.~B~138,~1~(1978)}.
%
%%CITATION = HEP-TH/0501015;%%
\bibitem{Kapustin:2005py}
A.~Kapustin,
\textit{``{Wilson-'t Hooft operators in four-dimensional gauge theories and
  S-duality}''},
\textsf{Phys.~Rev.~D74,~025005~(2006)},
\href{http://arXiv.org/abs/hep-th/0501015}{\texttt{hep-th/0501015}}.
%
\bibitem{Dirac:1931kp}
P.~A.~M.~Dirac,
\textit{``{Quantised singularities in the electromagnetic field,}''},
\textsf{Proc.~Roy.~Soc.~Lond.~A~133,~60~(1931)}.
%
\bibitem{Hull:2024uwz}
C.~M.~Hull,
\textit{``{Monopoles, Dirac Strings and Generalised Symmetries}''},
\href{http://arXiv.org/abs/2411.18741}{\texttt{2411.18741}}.
%
%%CITATION = NUPHA,B121,77;%%
\bibitem{Brink:1976bc}
L.~Brink, J.~H.~Schwarz and J.~Scherk,
\textit{``{Supersymmetric Yang-Mills Theories}''},
\textsf{Nucl.Phys.~B121,~77~(1977)}.
%
\bibitem{Stone:2020vva}
M.~Stone,
\textit{``{Gamma matrices, Majorana fermions, and discrete symmetries in
  Minkowski and Euclidean signature}''},
\textsf{J.~Phys.~A~55,~205401~(2022)},
\href{http://arXiv.org/abs/2009.00518}{\texttt{2009.00518}}.
%
\bibitem{Andrei:1980fv}
N.~Andrei,
\textit{``{Diagonalization of the Kondo Hamiltonian}''},
\textsf{Phys.~Rev.~Lett.~45,~379~(1980)}.
%
\bibitem{vigman1980exact}
P.~Wiegmann,
\textit{``Exact solution of sd exchange model at T= 0''},
\textsf{JETP~Lett.~31,~364~(1980)}.
%
\bibitem{wiegmann1981exact}
P.~Wiegmann,
\textit{``Exact solution of the sd exchange model (Kondo problem)''},
\textsf{J.~Phys.~C~14,~1463~(1981)}.
%
\bibitem{Tamm:1931dda}
I.~Tamm,
\textit{``{Die verallgemeinerten Kugelfunktionen und die Wellenfunktionen eines
  Elektrons im Felde eines Magnetpoles}''},
\textsf{Z.~Phys.~71,~141~(1931)}.
%
\bibitem{Fierz:1944}
M.~Fierz,
\textit{``{Zur Theorie magnetisch geladener Teilchen}''},
\textsf{Helvetica~Physica~Acta~17,~27~(1944)}.
%
\bibitem{Wu:1976qk}
T.~T.~Wu and C.~N.~Yang,
\textit{``{Dirac's Monopole Without Strings: Classical Lagrangian Theory}''},
\textsf{Phys.~Rev.~D~14,~437~(1976)}.
%
\bibitem{Wilczek:1981du}
F.~Wilczek,
\textit{``{Magnetic Flux, Angular Momentum, and Statistics}''},
\textsf{Phys.~Rev.~Lett.~48,~1144~(1982)}.
%
\bibitem{Wu:1977qk}
T.~T.~Wu and C.~N.~Yang,
\textit{``{Some Properties of Monopole Harmonics}''},
\textsf{Phys.~Rev.~D~16,~1018~(1977)}.
%
\bibitem{Weinberg:1993sg}
E.~J.~Weinberg,
\textit{``{Monopole vector spherical harmonics}''},
\textsf{Phys.~Rev.~D~49,~1086~(1994)},
\href{http://arXiv.org/abs/hep-th/9308054}{\texttt{hep-th/9308054}}.
%
\bibitem{Kazama:1976fm}
Y.~Kazama, C.~N.~Yang and A.~S.~Goldhaber,
\textit{``{Scattering of a Dirac Particle with Charge Ze by a Fixed Magnetic
  Monopole}''},
\textsf{Phys.~Rev.~D~15,~2287~(1977)}.
%
\bibitem{Olsen:1990jm}
H.~A.~Olsen, P.~Osland and T.~T.~Wu,
\textit{``{On the Existence of Bound States for a Massive Spin 1 Particle and a
  Magnetic Monopole}''},
\textsf{Phys.~Rev.~D~42,~665~(1990)}.
%
\bibitem{Rubakov:1981rg}
V.~A.~Rubakov,
\textit{``{Superheavy Magnetic Monopoles and Proton Decay}''},
\textsf{JETP~Lett.~33,~644~(1981)}.
%
\bibitem{Callan:1982ah}
C.~G.~Callan,~Jr.,
\textit{``{Disappearing Dyons}''},
\textsf{Phys.~Rev.~D~25,~2141~(1982)}.
%
\bibitem{Rubakov:1982fp}
V.~A.~Rubakov,
\textit{``{Adler-Bell-Jackiw Anomaly and Fermion Number Breaking in the
  Presence of a Magnetic Monopole}''},
\textsf{Nucl.~Phys.~B~203,~311~(1982)}.
%
\bibitem{Callan:1982au}
C.~G.~Callan,~Jr.,
\textit{``{Dyon-Fermion Dynamics}''},
\textsf{Phys.~Rev.~D~26,~2058~(1982)}.
%
\bibitem{Callan:1982ac}
C.~G.~Callan,~Jr.,
\textit{``{Monopole Catalysis of Baryon Decay}''},
\textsf{Nucl.~Phys.~B~212,~391~(1983)}.
%
\bibitem{Brennan:2021ewu}
T.~D.~Brennan,
\textit{``{Callan-Rubakov effect and higher charge monopoles}''},
\textsf{JHEP~2302,~159~(2023)},
\href{http://arXiv.org/abs/2109.11207}{\texttt{2109.11207}}.
%
\bibitem{Aharony:2023amq}
O.~Aharony, G.~Cuomo, Z.~Komargodski, M.~Mezei and A.~Raviv-Moshe,
\textit{``{Phases of Wilson lines: conformality and screening}''},
\textsf{JHEP~2312,~183~(2023)},
\href{http://arXiv.org/abs/2310.00045}{\texttt{2310.00045}}.
%
\bibitem{Bechler:1993}
A.~Bechler,
\textit{``{Summation formulae for spherical spinors}''},
\textsf{J.~Phys.~A:~Math.~Gen.~26,~6039~(1993)}.
%
%%CITATION = HEP-TH/0106015;%%
\bibitem{Semenoff:2001xp}
G.~W.~Semenoff and K.~Zarembo,
\textit{``{More exact predictions of SUSYM for string theory}''},
\textsf{Nucl.~Phys.~B616,~34~(2001)},
\href{http://arXiv.org/abs/hep-th/0106015}{\texttt{hep-th/0106015}}.
%
%%CITATION = HEP-TH/0604209;%%
\bibitem{Okuyama:2006jc}
K.~Okuyama and G.~W.~Semenoff,
\textit{``{Wilson loops in ${\cal N} = 4$ SYM and fermion droplets}''},
\textsf{JHEP~0606,~057~(2006)},
\href{http://arXiv.org/abs/hep-th/0604209}{\texttt{hep-th/0604209}}.
%
\bibitem{Gomis:2011pf}
J.~Gomis, T.~Okuda and V.~Pestun,
\textit{``{Exact Results for 't Hooft Loops in Gauge Theories on $S^4$}''},
\textsf{JHEP~1205,~141~(2012)},
\href{http://arXiv.org/abs/1105.2568}{\texttt{1105.2568}}.
%
%%CITATION = ARXIV:0904.4486;%%
\bibitem{Gomis:2009ir}
J.~Gomis, T.~Okuda and D.~Trancanelli,
\textit{``{Quantum 't Hooft operators and S-duality in N=4 super
  Yang-Mills}''},
\textsf{Adv.Theor.Math.Phys.~13,~1941~(2009)},
\href{http://arXiv.org/abs/0904.4486}{\texttt{0904.4486}}.
%
\bibitem{Gomis:2009xg}
J.~Gomis and T.~Okuda,
\textit{``{S-duality, 't Hooft operators and the operator product
  expansion}''},
\textsf{JHEP~0909,~072~(2009)},
\href{http://arXiv.org/abs/0906.3011}{\texttt{0906.3011}}.
%
%%CITATION = HEP-TH/0612073;%%
\bibitem{Gukov:2006jk}
S.~Gukov and E.~Witten,
\textit{``{Gauge Theory, Ramification, And The Geometric Langlands Program}''},
\href{http://arXiv.org/abs/hep-th/0612073}{\texttt{hep-th/0612073}}.
%
%%CITATION = ARXIV:0805.4199;%%
\bibitem{Drukker:2008wr}
N.~Drukker, J.~Gomis and S.~Matsuura,
\textit{``{Probing N=4 SYM With Surface Operators}''},
\textsf{JHEP~0810,~048~(2008)},
\href{http://arXiv.org/abs/0805.4199}{\texttt{0805.4199}}.
%
%%CITATION = HEP-TH/0010274;%%
\bibitem{Drukker:2000rr}
N.~Drukker and D.~J.~Gross,
\textit{``{An exact prediction of N = 4 SUSYM theory for string theory}''},
\textsf{J.~Math.~Phys.~42,~2896~(2001)},
\href{http://arXiv.org/abs/hep-th/0010274}{\texttt{hep-th/0010274}}.
%
%%CITATION = HEP-TH 0504190;%%
\bibitem{Beisert:2005fw}
N.~Beisert and M.~Staudacher,
\textit{``Long-range $PSU(2,2|4)$ Bethe ansaetze for gauge theory and
  strings''},
\textsf{Nucl.~Phys.~B727,~1~(2005)},
\href{http://arXiv.org/abs/hep-th/0504190}{\texttt{hep-th/0504190}}.
%
%%CITATION = ARXIV:1305.1939;%%
\bibitem{Gromov:2013pga}
N.~Gromov, V.~Kazakov, S.~Leurent and D.~Volin,
\textit{``{Quantum Spectral Curve for Planar $\mathcal{N} =$ Super-Yang-Mills
  Theory}''},
\textsf{Phys.~Rev.~Lett.~112,~011602~(2014)},
\href{http://arXiv.org/abs/1305.1939}{\texttt{1305.1939}}.
%
%%CITATION = HEP-TH/9911136;%%
\bibitem{Constable:1999ac}
N.~R.~Constable, R.~C.~Myers and O.~Tafjord,
\textit{``{The Noncommutative bion core}''},
\textsf{Phys.Rev.~D61,~106009~(2000)},
\href{http://arXiv.org/abs/hep-th/9911136}{\texttt{hep-th/9911136}}.
%
\bibitem{deLeeuw:2017dkd}
M.~de~Leeuw, A.~C.~Ipsen, C.~Kristjansen, K.~E.~Vardinghus and M.~Wilhelm,
\textit{``{Two-point functions in AdS/dCFT and the boundary conformal bootstrap
  equations}''},
\textsf{JHEP~1708,~020~(2017)},
\href{http://arXiv.org/abs/1705.03898}{\texttt{1705.03898}}.
%
\bibitem{Kristjansen:2020mhn}
C.~Kristjansen, D.~M\"uller and K.~Zarembo,
\textit{``{Integrable boundary states in D3-D5 dCFT: beyond scalars}''},
\textsf{JHEP~2008,~103~(2020)},
\href{http://arXiv.org/abs/2005.01392}{\texttt{2005.01392}}.
%
%%CITATION = HEP-TH/0111135;%%
\bibitem{DeWolfe:2001pq}
O.~DeWolfe, D.~Z.~Freedman and H.~Ooguri,
\textit{``{Holography and defect conformal field theories}''},
\textsf{Phys.Rev.~D66,~025009~(2002)},
\href{http://arXiv.org/abs/hep-th/0111135}{\texttt{hep-th/0111135}}.
%
%%CITATION = HEP-TH/0105132;%%
\bibitem{Karch:2000gx}
A.~Karch and L.~Randall,
\textit{``{Open and closed string interpretation of SUSY CFT's on branes with
  boundaries}''},
\textsf{JHEP~0106,~063~(2001)},
\href{http://arXiv.org/abs/hep-th/0105132}{\texttt{hep-th/0105132}}.
%
%%CITATION = ARXIV:1205.1674;%%
\bibitem{Nagasaki:2012re}
K.~Nagasaki and S.~Yamaguchi,
\textit{``{Expectation values of chiral primary operators in holographic
  interface CFT}''},
\textsf{Phys.Rev.~D86,~086004~(2012)},
\href{http://arXiv.org/abs/1205.1674}{\texttt{1205.1674}}.
%
%%CITATION = HEP-TH/9608163;%%
\bibitem{Diaconescu:1996rk}
D.-E.~Diaconescu,
\textit{``{D-branes, monopoles and Nahm equations}''},
\textsf{Nucl.Phys.~B503,~220~(1997)},
\href{http://arXiv.org/abs/hep-th/9608163}{\texttt{hep-th/9608163}}.
%
%%CITATION = ARXIV:1506.06958;%%
\bibitem{deLeeuw:2015hxa}
M.~de~Leeuw, C.~Kristjansen and K.~Zarembo,
\textit{``{One-point Functions in Defect CFT and Integrability}''},
\textsf{JHEP~1508,~098~(2015)},
\href{http://arXiv.org/abs/1506.06958}{\texttt{1506.06958}}.
%
%%CITATION = ARXIV:1512.02532;%%
\bibitem{Buhl-Mortensen:2015gfd}
I.~Buhl-Mortensen, M.~de~Leeuw, C.~Kristjansen and K.~Zarembo,
\textit{``{One-point Functions in AdS/dCFT from Matrix Product States}''},
\textsf{JHEP~1602,~052~(2016)},
\href{http://arXiv.org/abs/1512.02532}{\texttt{1512.02532}}.
%
\bibitem{Komatsu:2020sup}
S.~Komatsu and Y.~Wang,
\textit{``{Non-perturbative defect one-point functions in planar
  $\mathcal{N}=4$ super-Yang-Mills}''},
\textsf{Nucl.~Phys.~B~958,~115120~(2020)},
\href{http://arXiv.org/abs/2004.09514}{\texttt{2004.09514}}.
%
\bibitem{Beisert:2004ry}
N.~Beisert,
\textit{``{The Dilatation operator of N=4 super Yang-Mills theory and
  integrability}''},
\textsf{Phys.~Rept.~405,~1~(2004)},
\href{http://arXiv.org/abs/hep-th/0407277}{\texttt{hep-th/0407277}}.
%
\bibitem{Ivanovskiy:2024vel}
V.~Ivanovskiy, S.~Komatsu, V.~Mishnyakov, N.~Terziev, N.~Zaigraev and
  K.~Zarembo,
\textit{``{Vacuum Condensates on the Coulomb Branch}''},
\href{http://arXiv.org/abs/2405.19043}{\texttt{2405.19043}}.
%
%%CITATION = HEP-TH 0212208;%%
\bibitem{Minahan:2002ve}
J.~A.~Minahan and K.~Zarembo,
\textit{``The Bethe-ansatz for {$\mathcal{N}=\mathord{}$4} super Yang-Mills''},
\textsf{JHEP~0303,~013~(2003)},
\href{http://arXiv.org/abs/hep-th/0212208}{\texttt{hep-th/0212208}}.
%
%%CITATION = ARXIV:1203.1019;%%
\bibitem{Correa:2012nk}
D.~Correa, J.~Henn, J.~Maldacena and A.~Sever,
\textit{``{The cusp anomalous dimension at three loops and beyond}''},
\textsf{JHEP~1205,~098~(2012)},
\href{http://arXiv.org/abs/1203.1019}{\texttt{1203.1019}}.
%
%%CITATION = ARXIV:1803.02153;%%
\bibitem{Correa:2018lyl}
D.~H.~Correa, P.~Pisani and A.~Rios~Fukelman,
\textit{``{Ladder Limit for Correlators of Wilson Loops}''},
\textsf{JHEP~1805,~168~(2018)},
\href{http://arXiv.org/abs/1803.02153}{\texttt{1803.02153}}.
%
%%CITATION = HEP-TH/9911088;%%
\bibitem{Erickson:1999qv}
J.~Erickson, G.~Semenoff, R.~Szabo and K.~Zarembo,
\textit{``{Static potential in N=4 supersymmetric Yang-Mills theory}''},
\textsf{Phys.Rev.~D61,~105006~(2000)},
\href{http://arXiv.org/abs/hep-th/9911088}{\texttt{hep-th/9911088}}.
%
%%CITATION = HEP-TH/0602255;%%
\bibitem{Klebanov:2006jj}
I.~R.~Klebanov, J.~M.~Maldacena and C.~B.~Thorn,
\textit{``{Dynamics of flux tubes in large N gauge theories}''},
\textsf{JHEP~0604,~024~(2006)},
\href{http://arXiv.org/abs/hep-th/0602255}{\texttt{hep-th/0602255}}.
%
%%CITATION = ARXIV:1206.7117;%%
\bibitem{Bykov:2012sc}
D.~Bykov and K.~Zarembo,
\textit{``{Ladders for Wilson Loops Beyond Leading Order}''},
\textsf{JHEP~1209,~057~(2012)},
\href{http://arXiv.org/abs/1206.7117}{\texttt{1206.7117}}.
%
%%CITATION = ARXIV:1601.05679;%%
\bibitem{Gromov:2016rrp}
N.~Gromov and F.~Levkovich-Maslyuk,
\textit{``{Quark-anti-quark potential in $ \mathcal{N} =$ 4 SYM}''},
\textsf{JHEP~1612,~122~(2016)},
\href{http://arXiv.org/abs/1601.05679}{\texttt{1601.05679}}.
%
\bibitem{Beccaria:2016ejo}
M.~Beccaria, G.~Metafune and D.~Pallara,
\textit{``{The ground state of long-range Schr\"odinger equations and static
  $q\overline q$ potential}''},
\textsf{JHEP~1605,~040~(2016)},
\href{http://arXiv.org/abs/1603.03596}{\texttt{1603.03596}}.
%
\bibitem{Beccaria:2016bfi}
M.~Beccaria, A.~Fachechi and G.~Macorini,
\textit{``{On the cusp anomalous dimension in the ladder limit of $
  \mathcal{N}=4 $ SYM}''},
\textsf{JHEP~1606,~009~(2016)},
\href{http://arXiv.org/abs/1604.00897}{\texttt{1604.00897}}.
%
\bibitem{Cavaglia:2018lxi}
A.~Cavagli\`a, N.~Gromov and F.~Levkovich-Maslyuk,
\textit{``{Quantum spectral curve and structure constants in $ \mathcal{N}=4 $
  SYM: cusps in the ladder limit}''},
\textsf{JHEP~1810,~060~(2018)},
\href{http://arXiv.org/abs/1802.04237}{\texttt{1802.04237}}.
%
\bibitem{Correa:2018pfn}
D.~Correa, P.~Pisani, A.~Rios~Fukelman and K.~Zarembo,
\textit{``{Dyson equations for correlators of Wilson loops}''},
\textsf{JHEP~1812,~100~(2018)},
\href{http://arXiv.org/abs/1811.03552}{\texttt{1811.03552}}.
%
\bibitem{McGovern:2019sdd}
J.~McGovern,
\textit{``{Scalar insertions in cusped Wilson loops in the ladders limit of
  planar $ \mathcal{N} $ = 4 SYM}''},
\textsf{JHEP~2005,~062~(2020)},
\href{http://arXiv.org/abs/1912.00499}{\texttt{1912.00499}}.
%
\bibitem{Wadia:1980rb}
S.~R.~Wadia,
\textit{``{On the Dyson-schwinger Equations Approach to the Large $N$ Limit:
  Model Systems and String Representation of {Yang-Mills} Theory}''},
\textsf{Phys.~Rev.~D~24,~970~(1981)}.
%
\bibitem{Migdal:1983qrz}
A.~A.~Migdal,
\textit{``{Loop Equations and 1/N Expansion}''},
\textsf{Phys.~Rept.~102,~199~(1983)}.
%
\bibitem{Nagar:2024mjz}
I.~Nagar, A.~Sever and D.-l.~Zhong,
\textit{``{Planar RG flows on line defects}''},
\textsf{JHEP~2406,~110~(2024)},
\href{http://arXiv.org/abs/2404.07290}{\texttt{2404.07290}}.
%
%%CITATION = ARXIV:0909.4272;%%
\bibitem{Giombi:2009ek}
S.~Giombi and V.~Pestun,
\textit{``{The 1/2 BPS 't Hooft loops in N=4 SYM as instantons in 2d
  Yang-Mills}''},
\textsf{J.~Phys.~A46,~095402~(2013)},
\href{http://arXiv.org/abs/0909.4272}{\texttt{0909.4272}}.
%
%%CITATION = HEP-TH/0405001;%%
\bibitem{Beisert:2004hm}
N.~Beisert, V.~Dippel and M.~Staudacher,
\textit{``{A novel long range spin chain and planar N = 4 super Yang-
  Mills}''},
\textsf{JHEP~0407,~075~(2004)},
\href{http://arXiv.org/abs/hep-th/0405001}{\texttt{hep-th/0405001}}.
%
%%CITATION = HEP-TH/0511082;%%
\bibitem{Beisert:2005tm}
N.~Beisert,
\textit{``{The $su(2|2)$ dynamic S-matrix}''},
\textsf{Adv.~Theor.~Math.~Phys.~12,~945~(2008)},
\href{http://arXiv.org/abs/hep-th/0511082}{\texttt{hep-th/0511082}}.
%
%%CITATION = ARXIV:1309.1004;%%
\bibitem{Russo:2013kea}
J.~Russo and K.~Zarembo,
\textit{``{Massive N=2 Gauge Theories at Large N}''},
\textsf{JHEP~1311,~130~(2013)},
\href{http://arXiv.org/abs/1309.1004}{\texttt{1309.1004}}.
%
%%CITATION = HEP-TH/9803002;%%
\bibitem{Maldacena:1998im}
J.~M.~Maldacena,
\textit{``{Wilson loops in large N field theories}''},
\textsf{Phys.~Rev.~Lett.~80,~4859~(1998)},
\href{http://arXiv.org/abs/hep-th/9803002}{\texttt{hep-th/9803002}}.
%
\bibitem{Pufu:2013vpa}
S.~S.~Pufu,
\textit{``{Anomalous dimensions of monopole operators in three-dimensional
  quantum electrodynamics}''},
\textsf{Phys.~Rev.~D~89,~065016~(2014)},
\href{http://arXiv.org/abs/1303.6125}{\texttt{1303.6125}}.
%
\bibitem{Borokhov:2002ib}
V.~Borokhov, A.~Kapustin and X.-k.~Wu,
\textit{``{Topological disorder operators in three-dimensional conformal field
  theory}''},
\textsf{JHEP~0211,~049~(2002)},
\href{http://arXiv.org/abs/hep-th/0206054}{\texttt{hep-th/0206054}}.
%
\bibitem{Borokhov:2002cg}
V.~Borokhov, A.~Kapustin and X.-k.~Wu,
\textit{``{Monopole operators and mirror symmetry in three-dimensions}''},
\textsf{JHEP~0212,~044~(2002)},
\href{http://arXiv.org/abs/hep-th/0207074}{\texttt{hep-th/0207074}}.
%
\bibitem{Grignani:2007xz}
G.~Grignani, L.~Griguolo, N.~Mori and D.~Seminara,
\textit{``{Thermodynamics of theories with sixteen supercharges in non-trivial
  vacua}''},
\textsf{JHEP~0710,~068~(2007)},
\href{http://arXiv.org/abs/0707.0052}{\texttt{0707.0052}}.
%
\bibitem{gradshteyn2014table}
I.~Gradshteyn and I.~Ryzhik,
\textit{``Table of integrals, series, and products''},
Elsevier Science (2014).
%
\end{thebibliography}

\end{document}